\documentclass{aastex631}
\usepackage{appendix}
\pdfoutput=1
\usepackage{float}
\usepackage{amsmath,accents}
\usepackage{cases}
\usepackage{amssymb}
\usepackage{times}
\usepackage{graphicx}[pdftex]
\usepackage{epstopdf}
\usepackage{url}
\usepackage{multirow}
\usepackage{booktabs,bigstrut,multirow,rotating}
\usepackage{threeparttable}
\usepackage{multirow}
\usepackage{makecell}
\usepackage{xcolor}
\usepackage{bm}
\usepackage{ulem}
\def\r{\textcolor{red}}
\def\b{\textcolor{blue}}

\DeclareUnicodeCharacter{2212}{\textendash}

\received{XXX}
\revised{YYY}
\accepted{ZZZ}

\shorttitle{Exploring TeV candidates of \emph{Fermi} blazars}
\shortauthors{J. T. Zhu et al}

\begin{document}
\title{Exploring TeV candidates of \emph{Fermi} blazars through machine learning}

\correspondingauthor{J. H. Fan}
\email{fjh@gzhu.edu.cn}

\author[0000-0002-8206-5080]{J.~T.\ Zhu}
\affiliation{Center for Astrophysics, Guangzhou University, Guangzhou 510006, China}
\affiliation{Department of Physics and Astronomy ``G. Galilei'', University of Padova, Padova I-35131, Italy}
\affiliation{Astronomy Science and Technology Research Laboratory of Department of Education of Guangdong Province, Guangzhou 510006, China}
\affiliation{Key Laboratory for Astronomical Observation and Technology of Guangzhou, Guangzhou 510006, China}

\author{C.\ Lin}
\affiliation{Center for Astrophysics, Guangzhou University, Guangzhou 510006, China}
\affiliation{Astronomy Science and Technology Research Laboratory of Department of Education of Guangdong Province, Guangzhou 510006, China}
\affiliation{Key Laboratory for Astronomical Observation and Technology of Guangzhou, Guangzhou 510006, China}

\author{H.~B.\ Xiao}
\affiliation{Shanghai Key Lab for Astrophysics, Shanghai Normal University Shanghai, 200234, China}

\author[0000-0002-9071-5469]{J.~H.\ Fan}
\affiliation{Center for Astrophysics, Guangzhou University, Guangzhou 510006, China}
\affiliation{Astronomy Science and Technology Research Laboratory of Department of Education of Guangdong Province, Guangzhou 510006, China}
\affiliation{Key Laboratory for Astronomical Observation and Technology of Guangzhou, Guangzhou 510006, China}

\author[0000-0002-6954-8862]{D.\ Bastieri}
\affiliation{Center for Astrophysics, Guangzhou University, Guangzhou 510006, China}
\affiliation{Department of Physics and Astronomy ``G. Galilei'', University of Padova, Padova I-35131, Italy}
{\Large }

\author{G.~G.\ Wang}
\affiliation{Center for Astrophysics, Guangzhou University, Guangzhou 510006, China}
\affiliation{Astronomy Science and Technology Research Laboratory of Department of Education of Guangdong Province, Guangzhou 510006, China}
\affiliation{Key Laboratory for Astronomical Observation and Technology of Guangzhou, Guangzhou 510006, China}

\begin{abstract}

In this work, we make use of a supervised machine learning algorithm based on Logistic Regression (LR) to select TeV blazar candidates from the 4FGL-DR2 / 4LAC-DR2, 3FHL, 3HSP, and 2BIGB catalogs. LR constructs a hyperplane based on a selection of optimal parameters, named features, and hyper-parameters whose values control the learning process and determine the values of features that a learning algorithm ends up learning, to discriminate TeV blazars from non-TeV blazars. In addition, it gives the probability (or logistic) that a source may be considered as a TeV blazar candidate.
Non-TeV blazars with logistics greater than 80\% are considered high-confidence TeV candidates. Using this technique, we identify 40 high-confidence TeV candidates from the 4FGL-DR2 / 4LAC-DR2 blazars and we build the feature hyper-plane to distinguish TeV and non-TeV blazars. We also calculate the hyper-planes for the 3FHL, 3HSP, and 2BIGB. Finally, we construct the broadband spectral energy distributions (SED) for the 40 candidates, testing for their detectability with various instruments. We find that 7 of them are likely to be detected by existing or upcoming IACT observatories, while 1 could be observed with EAS particle detector arrays.

\end{abstract}


\section{Introduction}
Blazars, an extreme subclass of active galactic nuclei (AGNs), are known for prominent observation properties, such as high energy $\gamma$-ray emissions, rapid and significant amplitude variability, high luminosity, high and variable polarization, and superluminal motions, etc. \citep{Wills1992, Urry1995, Fan2002, Villata2006, Fan2014, Xiao2015, Gupta2016, Xiao2019, Abdollahi2020, Xiao2020, Fan2021}.
Blazars are historically subdivided into two main categories based on the equivalent width (EW) of the optical emission lines: flat spectrum radio quasars (FSRQs) and BL Lacertae objects (BL Lacs).
In general, FSRQs show an EW greater than 5 $\mathring{A}$, while BL Lacs illustrate no or weak emission lines, with EW less than 5 $\mathring{A}$.
Meanwhile, the spectral energy distributions (SEDs) of blazars are generally characterized by two well-separated bumps, a low-energy one is in the infrared to soft X-ray energy range due to synchrotron emission, and a high-energy one is in the region between hard X-ray to $\gamma$-ray that is associated with an inverse Compton (IC) radiation according to leptonic model (IC; e.g., \citealt{1994ApJ...421..153S}). The seed photons undergoing IC scattering could be from the same electron population producing the synchrotron bump in the so-called self-Compton (SSC) model \citep{1985A&A...146..204G, 1992ApJ...397L...5M, 1996ApJ...461..657B}, e.g. 1ES 0347-121, 1ES 0229+200 \citep{2018MNRAS.477.4257C, 2007A&A...473L..25A, 2007A&A...475L...9A}, or from external regions (External Compton model, EC), e.g., from the accretion disk \citep{disk}, broad line region \citep{BLR}, and dust torus \citep{torus}. While the hadronic process could also contribute to the high-energy bump
by high-energy cosmic rays via the photohadronic reactions \citep{2012A&A...546A.120D, 10.1093/mnras/stt2437}. SEDs are also used to make classifications of blazars.
In the original BL Lac classification of \citet{Padovani1995}, BL Lacs are classified as two subclasses: `low synchrotron peaked (LSP) BL Lac' or LBL and `high synchrotron peaked sources (HSP) BL Lac' or HBL, depending on their broadband radio-to-X-ray spectral index is larger or smaller than 0.75. \citet{Abdo2010} later extended the classification to all blazars. They suggested to group blazars into three subclasses based on synchrotron peak frequency, LSP, $\log \nu_{\rm p}^{\rm s} \leq 14.0$; ISP (intermediate synchrotron peaked sources), $14 < \log \nu_{\rm p}^{\rm s} \leq 15$; and HSP, $\log \nu_{\rm p}^{\rm s} > 15$. While, \citet{Fan2016} proposed slightly different criteria: $14 < \log \nu_{\rm p}^{\rm s} \leq 15.3$ for ISP, and $\log \nu_{\rm p}^{\rm s} > 15.3$ for HSP. Very recently, \cite{2022ApJS..262...18Y} also gave a similar classification.

Besides, \citet{Costamante2001} proposed a fourth subclass namely extreme HBLs (EHBLs) for BL Lacs having $\log \nu_{\rm p}^{\rm s} > 17$.
The EHBLs can be divided into three subclasses according to their IC bump peak \citep{2019MNRAS.486.1741F} with the subclass with IC bump peak between $0.1\:\mathrm{TeV}$ and $1\:\mathrm{TeV}$ being a continuation of HSP, while the subclass with IC bump peak $>10\:\mathrm{TeV}$ named  `hard-TeV blazars' is an independent category featuring high power, very stable flux, and hard-TeV spectral behavior at TeV energies, posing a challenge to the SSC.
With an IC bump peaking at a few TeV, the remaining subclass behaves as a transition class with a flat TeV spectral slope.
Thus, the HSPs and HBLs are often considered as the potential emitters of very high energy (VHE, above 300 GeV) radiations, among which, the sources with small redshift have a significant fraction of TeV photons to be observed. The emissions of VHE photons with energy $>300\:\mathrm{GeV}$ from blazars reveals new phenomena, especially photons in the TeV band raise the challenge of particle acceleration in jets, and they are also essential clues for indirectly measuring the extragalactic background light, estimating the intergalactic magnetic field, and probing the possible origin of high-energy extragalactic neutrinos \citep{2019MNRAS.486.1741F}. For instance, the Very Long Baseline Array (VLBA) provided 23 images of 6 TeV blazars and proved that apparent jet bending is a common property of TeV blazars \citep{2010ApJ...723.1150P}.

The emitted flux from most astronomical sources is very low in the TeV band. In addition, the extragalactic background light (EBL) absorbs the most energetic $\gamma$-ray emissions \citep{ebl1, ebl2, 2018Sci...362.1031F} via the interaction $\gamma$ + $\gamma$ $\rightarrow$ $e^+$ + $e^-$, which is relative to many fundamental astrophysics problems \citep{Dom_nguez_2019}.
Therefore, observations at TeV energies require large collection areas, which are only affordable for ground-based detectors like atmospheric Cherenkov telescopes (IACTs) and extended arrays of particle detectors (EAS arrays).
The $\gamma$-rays detection techniques of IACTs and EAS arrays are different. IACTs detect Cherenkov photons in the atmosphere generated by the atmospheric extended shower (EAS) of the secondary particles initiated by the primary $\gamma$-rays, and popular examples are 
\emph{the High Energy Stereoscopic System}
\footnote{https://www.mpi-hd.mpg.de/hfm/HESS/}
(\citealt{HESS}, H.E.S.S),
\emph{the Major Atmospheric Gamma Imaging Cherenkov Telescopes}
\footnote{https://magic.mpp.mpg.de/}
(\citealt{MAGIC}, MAGIC), 
\emph{the Very Energetic Radiation Imaging Telescope Array System}
\footnote{https://veritas.sao.arizona.edu/} 
(\citealt{VERITAS}, VERITAS) and the next-generation IACT array: \emph{the Cherenkov Telescope Array Observatory} 
\footnote{https://www.cta-observatory.org/}
(\citealt{CTAO}, CTAO); While EAS arrays detect
secondary particles from the cosmic ray surviving down to the ground, such as 
\emph{the High Altitude Water Cherenkov Observatory}
\footnote{https://www.hawc-observatory.org/}
(\citealt{HAWC}, HAWC), and
\emph{the Large High Altitude Air Shower Observatory}
\footnote{http://english.ihep.cas.cn/lhaaso/}
(\citealt{LHAASO}, LHAASO).

Up to August 2022, only about 90 extragalactic TeV sources \citep{TeVCat}\footnote{http://TeVCat.uchicago.edu}, out of which are 81 TeV blazars, have been verified by IACTs and EAS arrays. Finding TeV blazar candidates is an exciting and challenging work since the EBL and the sensitivity of the detector limits the number of TeV sources with high redshift. Therefore, under the constraints of EBL, an improved sensitivity will expand the sample of TeV blazars in terms of the sheer number and will find sources with higher redshift. Shortly, current and next-generation of IACTs and EAS arrays are sensitive to photons at TeV energies should increase the TeV blazar population, helping us to explore this classification further and understand the mechanism of VHE emission. \citep{2002AA...384...56C,
2013ApJS..207...16M,
2017AA...598A..17C,
2019MNRAS.486.1741F}.
The current TeV $\gamma$-ray detection is offered by the large area of ground-based detectors.

Most of the TeV blazars are HSPs (almost all BL Lacs),
thus the TeV blazar candidate searches are often limited to HSPs/BL Lacs. (e.g. \citealt{2002AA...384...56C, 2013ApJS..207...16M, 2017AA...598A..17C, 2019MNRAS.486.1741F, 2019ApJ...887..104C}). Typically, BL Lacs are HSPs, whereas FSRQs are mostly LSPs and ISPs. LSP/ISP often have EC components where the electrons see a strong photon field that is Doppler-shifted. In principle they are capable of producing strong $\gamma$-ray emissions, sometimes entering the VHE regime. But in the case of LSP/ISPs, the interaction often happens in the Klein-Nishina regime and the same strong photon field often induces a very strong absorption of the produced $\gamma$-ray photons.
Compared to LSP/ISPs, the VHE emission of HSPs is thought to originate from the low-energy photons produced by synchrotron radiation by ultrarelativistic electrons in the jet according to the SSC process. Besides, HSPs have higher $\log \nu_{\rm p}^{\rm s}$, making their synchrotron bump more easily scattered into the TeV band by the relativistic electrons.
However, it is not obvious that extreme $\log \nu_{\rm p}^{\rm s}$ values lead to the emission of TeV $\gamma$-rays by itself, as it also depends on many other properties of the emission region, such as electron distribution, magnetic field strength, internal absorption, and the redshift of the source \citep{2022MNRAS.512..137N}.

\citet{Lin2016} gathered 662 Fermi BL Lacs, including 47 TeV sources, and compared the multiwavelength observation properties of the TeVs with those of the non-TeVs. They discovered that TeVs have a smaller average redshift, a higher flux density, and a harder $\gamma$-ray spectrum.
Flux variability in all wavebands is very common among blazars \citep{2017ApJ...837...45F, 2019MNRAS.490..124M, 2022ApJS..262...18Y, 2022RAA....22h5002Y}, with some blazars, such as 1ES 0229+200, exhibiting moderate variability and others displaying violent and high amplitude variability, with timescales ranging from hours to years.
In addition, the $\log \nu_{\rm p}^{\rm s}$ is proportional to the variability of blazars.
\citet{gupta2016peak} analyzed 50 observations of 12 low synchrotron peaked (LSP) blazars from \emph{XMM–Newton} and discovered that LSP blazars vary more slowly in the X-ray bands than in the IR/optical bands because the IC mechanism dominates the X-ray radiation in this case.
In contrast, HSP blazars are predicted to exhibit more extreme X-ray band variability than LSP blazars.
According to the SSC model, relativistic electrons upscatter X-rays into the TeV region, and
observations have confirmed correlations between X-ray and TeV emissions \citep{2002AA...384...56C, 2019AdSpR..63..766S, 2019ICRC...36..686O}.
One can therefore anticipate a correlation between X-ray and TeV emissions. Notably, Markarian 501 and 1ES 1959+650's synchrotron peak frequencies in the X-ray band shifted to the higher energy region when a flare was observed in the TeV band \citep{2000ApJ...538..127S, 2018MNRAS.473.2542K, 2019AdSpR..63..766S}.
In contrast, there exists an intriguing counterexample.
\citet{2019MNRAS.486.1741F} observed two groups of EHBLs with the same range of $\log \nu_{\rm p}^{\rm s}$ but opposite spectral slope in the TeV band, implying that $\log \nu_{\rm p}^{\rm s}$ and the TeV emission are independent of one another.

Beyond all doubt, the \emph{Fermi} Large Area Telescope (\emph{Fermi}-LAT) brought prosperity to the study of blazars after its launch in 2008, and the progress of high-energy $\gamma$-ray astronomy of space-borne instruments is gradually dominated by it.
\emph{Fermi}-LAT detected thousands of blazars and published them in their works (e.g.,\citealt{4FGL}), which provided a large $\gamma$-ray blazar sample. The \emph{Fermi-LAT 10-year source catalog} (4FGL-DR2, \citealt{4FGL, 4FGL-DR2}), and the \emph{Fermi}-LAT 10-year AGN catalog (4LAC-DR2, \citealt{4LAC, 4LAC-DR2}) listed 80 out of the 81 TeV blazars (hereafter \textit{Fermi} blazars.)
Thus, we have a convincing reason to believe that \textit{Fermi} catalogs should contain many TeV blazar candidates.

Machine learning (ML) techniques have become popular among astronomers \citep{ML_and_DM, 2019ApJ...887..104C, Kang2019, Xiao2022}.
Based on the \emph{Fermi}-LAT third source catalog (3FGL), \citet{2019ApJ...887..104C} applied the Artificial Neural Network (ANN) to 573 uncertain types and 559 unassociated catalog sources of the 3FGL blazars and found 80 HSPs candidates, 16 of them are proposed as TeV candidates with the highest confidence.
\citet{Kang2019} collected 1312 blazar candidates of uncertain type (BCUs) out of 3137 blazars recorded in \emph{Fermi}-LAT 4th source catalog, then employed random forest (RF), support vector machine (SVM), and ANN to predict the category of the BCUs. Finally, 724 BL Lac candidates and 332 FSRQ candidates are predicted by combined classification results of the ML methods.
\citet{Xiao2022} compiled radio loudness ${\rm log} R$ and radio 6 cm  luminosity ${\rm log} L_{6 cm}$ of 2943 AGNs, and got ${\rm log} R = \langle 1.37 \pm 0.02 \rangle$ as the separation of radio-loud and radio-quiet AGNs by the Gaussian Mixture Model.
Furthermore, a more advanced double-criterion dividing boundary ${\rm log} L_{\rm 6cm} = −2.7{\rm log} R + 44.3$ was obtained by the SVM.

From a more comprehensive perspective, this work aims to increase the number of TeV blazar candidates, and break through the limitation of searching for candidates only in HSPs. Using the machine learning method, we look for the physical properties that truly distinguish TeV from non-TeV blazars, and quantify the performance of distinguishing boundaries. We also calculate the probability that a source can be called a TeV source based on these physical properties, and find sources with a higher probability of being called a TeV source from non-TeV sources. The data are compiled from four catalogs: 4FGL-DR2 / 4LAC-DR2, the 3rd Catalog of Hard Fermi-LAT Sources (3FHL, \citealt{3FHL}), the 3rd catalog of HSP blazars (3HSP, \citealt{3HSP}), and The Second Brazil-ICRANet Gamma-ray Blazars (2BIGB, \citealt{2BIGB}). Then make sure how many candidates could be detected by the ground-based Cherenkov telescopes.
The paper is organized as follows:
Section 2 introduces the ML method;
Section 3 gives the experiments and results;
We make discussion and draw the conclusion in Section 4.

\section{Supervised machine learning method}
We intend to use the so-called supervised machine learning (SML) to select TeV blazar candidates among four blazar catalogs.
In ML, samples are often referred to as datasets. 
SML considers a labeled dataset as a set of features and a target variable, based on example input-output pairs.
The SML method constructs an inferred function (often called a model) to map the features to dispersed target variables from the known dataset in a classification task. Moreover, the model can predict labels for the unknown dataset. The dataset will be divided into three sets: a training set acts as known data to train the model, which is further decomposed into a smaller training set and a validation set which is used to improve the model, and, finally, the test set, acts as unclassified data to evaluate the generalization of the model. The dispersed target variables are often called labels, which mark the class of data \citep{ML_and_DM, ML}.
Each source catalog can be regarded as a dataset in a matrix. The parameter columns are the features that characterize the source physical properties.
Besides, binary labels distinguish TeV blazars from non-TeV blazars.
Our work gets the model from training sets and then employs them in the whole dataset. Consequently, the non-TeV sources with the same predicted labels as the TeV sources are considered TeV blazar candidates. It is taken as a high-confidence candidate if the possibility of being a TeV source is higher than 80\%.

\begin{figure*}[htbp]
\centering
\includegraphics[width = 4 in]{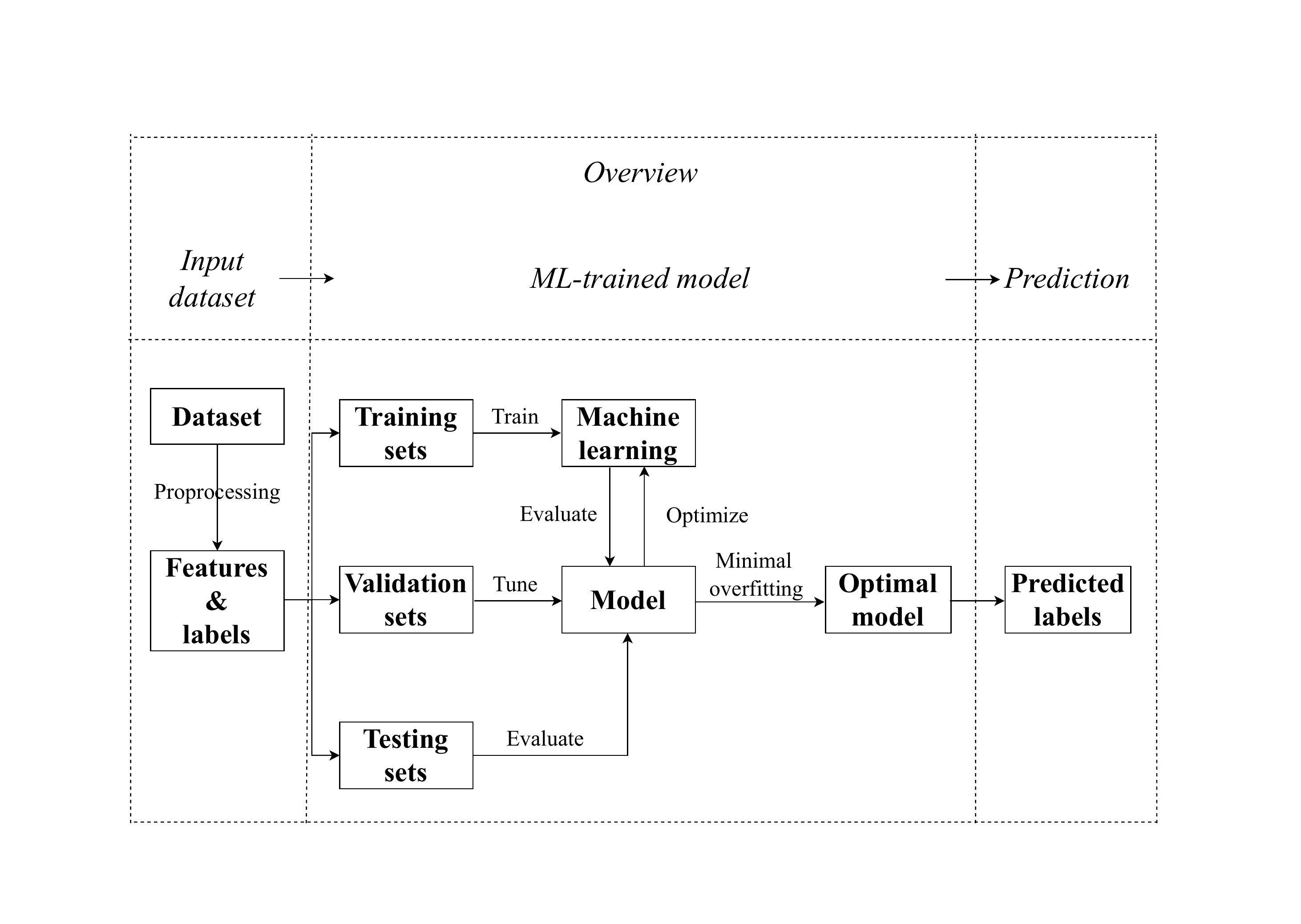}
\caption{Flowchart of SML.}
\label{ml_flowchart}
\end{figure*}

There are several steps to accomplish this goal, which is demonstrated in Fig.~\ref{ml_flowchart}.
The process of SML classification can be roughly divided into three parts: \emph{Input dataset, ML-trained model, and prediction}. Next, we dive into details.

\subsection{Data preprocessing}\label{preprocess}
Data preprocessing is an essential step in the dataset before starting the training model.
By manipulating or deleting the data, preprocessing of data can change natural features into a more available representation for the downstream model. It contains the following steps:

\begin{description}
\item[Data Discretization]
In some of the datasets, one feature is represented by multiple columns.
For example, 
in 4FGL-DR2, such as $Flux\_band$, 
7 columns represent integral photon flux in each spectral band.
We divide $Flux\_band$ as $Flux\_band\ 1$, $Flux\ band\ 2$, ... $Flux\_band\ 7$, respectively.
\item[Data cleaning]
There may be some \emph{useless} and missing data, which will reduce model performance. We discard error features, string features, and most missing value features.
\item[Data standardization]
To make all features dimensionless,
we use \emph{sklearn.preprocessing.StandardScaler} to convert all features into the standard normal distribution.
\item[Data partition]
We randomly divide the dataset into a 4:1 training set and testing set using \textsf{python} package \emph{StratifiedKFold.split} of \emph{sklearn.model\_selection}, and each training / testing set preserves the percentage of the dataset for each class. We repeat the division 5 times with 5 different random seeds: 0, 1, 2, 3, 4 to get 5 training sets and 5 testing sets. We mark the training / testing sets as training 1 / testing 1, training 2 / testing 2, ..., training 5 / testing 5.
\end{description}

\subsection{Selecting SML model}
We chose the SML method based on the following considerations:
\begin{enumerate}
\item We request a highly interpretable model that is not overly dependent on computer performance, which outputs an explicitly linear boundary to separate TeV blazars and non-TeV blazars in the feature space.
\item Once we have the physical properties of a blazar, that are, the features, we want to get the conditional probability that the blazar is a TeV one under the features.
\end{enumerate}

There are many highly-developed SML classification methods, such as:
\emph{Logistic Regression} (\citealt{logistic_regression}, LR)
\emph{Support Vector Machines} (\citealt{SVM}, SVM),
\emph{artificial neural network} (\citealt{ann}, ANN), 
\emph{random forest} \citep{RF}, etc.
In this work, we choose the LR model over other methods like SVM, ANN and the likes for two main reasons. Mathematically, we assumed that the binary labels of blazars follow a Bernoulli distribution, and, as the logistic function is the expectation of a Bernoulli distribution, we believe it is the most natural way to map the linear combination of physical properties of the source into a Bernoulli distribution. Algorithmically, SVM, RF, and ANN are more complicated than LR, resulting in higher calculation costs and an overall loss in the ability to interpret the model, moreover the output of SVM needs to go through the LR model to obtain a conditional probability. The same holds true for ANN, although other models may be used to obtain conditional probability. In addition, ANN and RF cannot guarantee a linear boundary.

To illustrate LR,
we introduce the odds ratio: $\frac{p}{1-p}$,
where $p$ represents the probability of the positive event labeled 1.
Then we further define the logit function (log-odds),
which can be written as ${\rm logit} \, (p) = {\rm log}\, \frac{p}{1-p}$.
We can express a linear relationship between features and the log-odds:
\begin{equation}
\label{logit_equa}
logit\, (p(y = 1 | \mathbf{x}))=w_{0} + w_{1} x _{1} + \cdots + w_{m} x _{m} = \mathbf{w}^T\mathbf{x},
\end{equation}
here $y$ is the label; 
the positive event is marked as `1',
while `0' for the negative events.
$p (y = 1|\mathbf{x}$) is the conditional probability that an individual labeled as `1',
and $\mathbf{w}$ is the weight.
$p (y = 1|\mathbf{x}$) (or $logistic$) is the inverse function of the ${\rm logit}$ function called the ${\rm logistic}$ function (or sigmoid function),
, which can be expressed as:

\begin{equation}
\label{logistic_equa}
logistic\, (logit (p)) = \frac{1}{1 + e^{- logit (p)}},
\end{equation}
where $logit (p)$ means $logit\, (p(y=1|\mathbf{x}))$.

For the implementation, we used \textit{scikit-learn} (sklearn, \citealt{sklearn}), an ML library in \textsf{python}, which provides many useful tools.
In particular, we used a sklearn API
\citep{sklearn-api}:
\emph{sklearn.linear\_model.LogisticRegression} which implements the LR model,
to train the training set.
Eventually,
we used the LR to classify the dataset by evaluating the $logit$ value of each source in the dataset and calculating the conditional probability $logistic$ that a source emits TeV radiation. Note that the number of TeV sources in our sample is small.
To increase the weight of TeV sources, we set the weight parameter \emph{class\_weight} to \emph{balance} in \emph{sklearn.linear\_model.LogisticRegression}.

\subsection{Evaluating performance}
\begin{figure*}[htbp]
\centering
\includegraphics[width = 1.5 in]{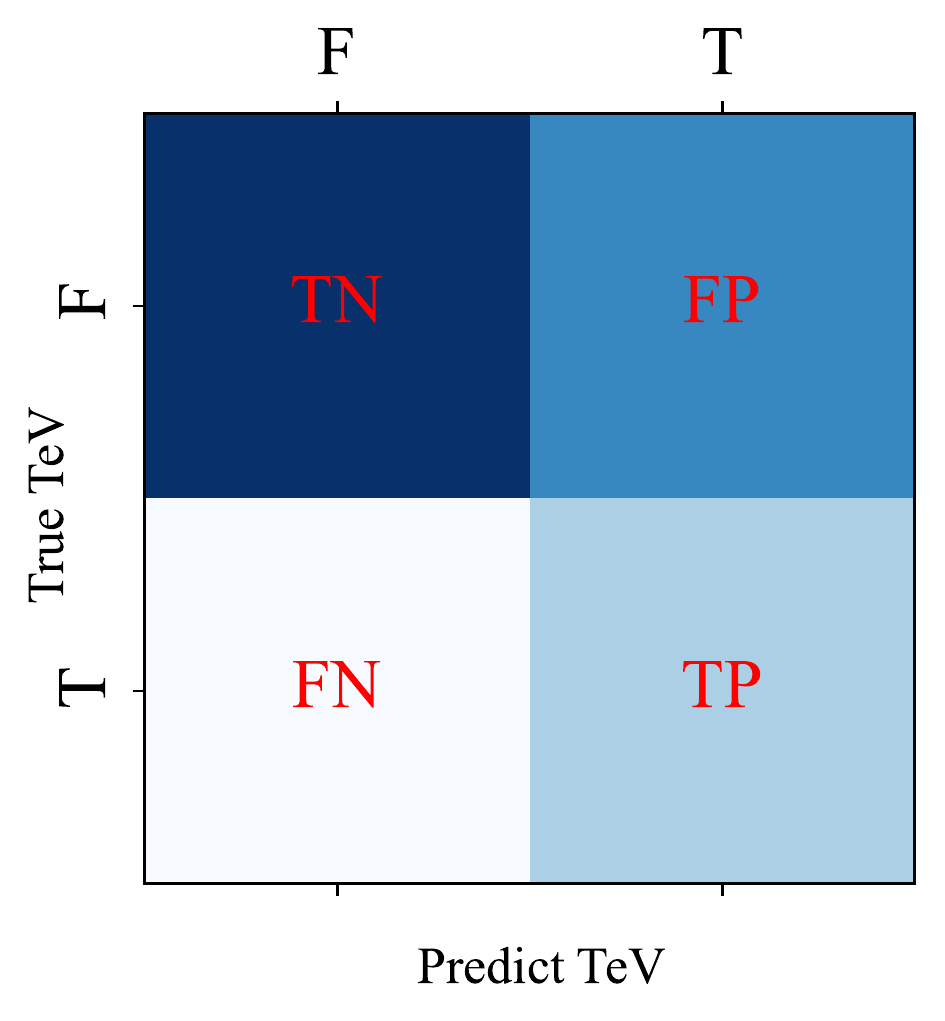}
\caption{Plot of the confusion matrix, where the coordinate value `T' stands for `True', and `F' stands for `False'.}
\label{plotcmat}
\end{figure*}

A confusion matrix is applied to visualize the model performance.
True positive (TP) represents the number of positive events that are correctly classified,
true negative (TN) represents the number of negative events that are correctly classified,
false positive (FP) represents the number of misclassified positive events,
false negative (FN) represents the number of misclassified negative events,
as shown in Fig.~\ref{plotcmat}.

The accuracy ($\frac{TP + TN}{FP + FN + TP + TN}$) is one of the most common indexes for evaluating model performance.
However, we have an extreme imbalance of the two classes,
which means non-TeV sources dominate the results.
We also want the model to identify as many TeV sources as possible while reducing the number of non-TeVs misjudged as TeVs.
To meet these two requirements, we introduce \textit{larger Area Under the Curve} (\citealt{ROC}, $AUC$) as the metric to evaluate LR performance, which is based on the \textit{receiver operating characteristic} curve (\citealt{ROC}, \textit{ROC}) with the true positive rate ($TPR$) on the Y-axis and the 1- false positive rate ($FPR$) on the X-axis, where $TPR = \frac{TP}{TP + FN}$ and $FPR = \frac{FP}{FP + TN}$. Furthermore, we introduce Youden's J statistic (or Youden's index, \citealt{Youden_index}), which is $TPR - FPR$.
The optimal $p_{\rm thre}$ corresponds to the maximum Youden index, which means that when we fix TPR, FPR reaches a minimum, which indicates that our screening of TeV candidates is very stringent.
So we consider a non-TeV blazar as TeV one if its $logistic$ beyond the optimal $p_{\rm thre}$.

\subsection{Training and tuning model}
Now we can train the models on the 5 training sets and further optimize the model performance in two ways:
\begin{description}
\item [Feature selection]
It may degrade model performance if all features are used,
which is called the \textit{curse of dimensionality} \citep{curse}. Data becomes sparse in high-dimensional space, resulting in no more extended similarities between data, which hinders efficient organization and data processing.

The \emph{sklearn.feature\_selection.SequentialFeatureSelector} (or SFS for short) of \textsf{python} could filter useless features, where the basic idea is to sequentially remove features from the full feature space until the new feature subspace contains the desired number of features.

\item [Hyper-parameters optimization]
Some parameters in the model cannot be obtained from the training model, which is often called hyper-parameters; they are not the $\mathbf{w}$ in Formula~(\ref{logit_equa}). Instead, we need to specify them manually before the training model.

The \emph{sklearn.grid\_search.GridSearchCV} (or GS for short) is helpful in the selection of optimal hyper-parameters, which performs a brute-force search on a list of values that we specify for hyper-parameters, and picks out the optimal hyper-parameters.
For \emph{sklearn.linear\_model.LogisticRegression}, we optimize 6 hyper-parameters \footnote{https://scikit-learn.org/stable/modules/generated/sklearn.linear\_model.LogisticRegression.html}: \emph{penalty, C, solver, tol, max\_iter, l1\_ratio}.
\end{description}

The optimization strategy is inspired by the \emph{k-fold cross-validation} \citep{ML_and_DM},
which randomly divides the training set into \emph{k} folds without replacement. Then \emph{k - 1} folds are used for the training model, and the remaining one fold, namely the validation set, evaluates the model performance. This process was repeated \emph{k} times.

Based on \emph{k-fold cross-validation}, we do both feature selection and hyper-parameter optimization at the same time using the \emph{nested cross-validation} \citep{PYML, nested}, which is characterized by a nested loop. The dataset in the outer loop is divided into a training set and a testing set. The training set in the inner loop undergoes k-fold cross-validation selecting optimal hyper-parameters, which is performed by GS. Next, the SFS in the outer loop picks out the optimal features,and the testing set evaluates the model generalization.
We implement nested cross-validation on 5 partitions of the dataset and get 5 models with optimal features and hyper-parameters.

However, 
Bias-variance trade-off is one of the fundamental rules of machine learning.
The bias measures the deviation between the predicted value and the actual value on the known dataset,
and the variance measures the performance of the model on the unknown dataset.
Overfitting is a common problem in machine learning, and it is mainly due to high variance, making the performance on the testing set worse than that on the training set. 
Therefore, for each of the 5 partitions, we calculate the gap that equals $AUC$ of the training set minus $AUC$ of the testing set and selects one optimal model corresponding to the smallest positive gap.

\subsection{Constructing linear boundary}
Now we could build linear boundaries to distinguish TeV emitting sources from non-TeV emitting sources.
When the $logistic$ of a source is greater than a threshold value ($p_{\rm thre}$), the LR determines the source as a TeV source, otherwise as a non-TeV one. A $p_{\rm thre}$ corresponds a $logit$ threshold value: $logit_{thre}$ (See Formula~(\ref{logistic_equa})). So we could express the linear boundary as $logit^{\prime} = 0$, where $logit^{\prime} = logit - logit_{thre}$.

\subsection{Predicting labels}
To better understand the process of predicting labels for the unknown datasets, we define some transition concepts. According to section ~\ref{preprocess}, we call the dataset that has only undergone \textit{data discretization} as the \textit{initial set}. In contrast, the dataset that has undergone complete preprocessing is called the \textit{learning set}. We get the optimal model from the \textit{learning set}, and the optimal features are also determined. In the \textit{initial set}, a subset containing the optimal features is called a \textit{prediction set}. The \textit{learning set} requires that all features have no missing values, but the prediction set only needs the full optimal parameters, containing more samples that will help to find more TeV candidates. 
Consequently, we could calculate the $logistic$ for each blazar in the \textit{prediction set}, and consider the non-TeV blazars with $logistic \geqslant 80\%$ as high-confidence TeV candidates.

\section{LR model to distinguish TeVs from non-TeVs}
\subsection{Samples}
We used four catalogs: 4FGL-DR2 / 4LAC-DR2, 3FHL, 3HSP, and 2BIGB to complete our tasks.
4FGL-DR2, based on 10-year Fermi-LAT data, contains over 5700 sources, 3511 of which are AGNs (3436 are blazars).
In the observer frame, 4LAC-DR2 includes synchrotron peak frequency ($\log \nu_{\rm p}^{\rm s}$), the corresponding intensity ($\log \nu_{\rm p}^{\rm s}f_{\nu_{\rm p}}^{\rm s}$), and redshift ($z$), which are critical supplements to 4FGL-DR2; 3FHL \citep{3FHL} is a 7-year-based catalog of 1556 VHE sources;
3HSP
\citep{3HSP}
includes 2013 HSP blazars containing a critical feature: the peak flux in units of the faintest HSP blazar peak flux
(2.5$\times$ 10$^{-12}$ $\:\mathrm{erg \cdot cm^{-2} \cdot s^{-1}}$) that has been detected in the TeVCat\citep{TeVCat}: $FOM$ (Figure of Merit);
2BIGB
\citep{2BIGB}
is the result of $\gamma$-ray likelihood analysis for 1160 3HSP sources yielded photon flux from 500 MeV to 500 GeV: $\rm F_{0.5-500 GeV}$ ($F^{\rm ph}$).

TeVCat is an online interactive catalog containing VHE sources.
As of now, it contains 251 sources detected mainly by IACTs and EAS arrays in the TeV energy region,
and at least 91 are extragalactic sources. 
Among those sources, 
80 blazars are detected by \emph{Fermi}-LAT, thus we assume that blazars emitting TeV radiation are usually detected by \emph{Fermi}-LAT.
We cross-matched the four catalogs with TeVCat, respectively. We marked labels of their blazars in common with TeVCat as `1', while the others were` 0'. According to the SML workflow introduced in the previous section, 
we got four groups of \textit{initial, learning,} and \textit{prediction sets}:
\begin{description}
\item [4FGL-DR2 / 4LAC-DR2]
We supplement 4FGL-DR2 with $\log \nu_{\rm p}^{\rm s}$, $\log \nu_{\rm p}f_{\nu_{\rm p}}$ and $z$ from 4LAC-DR2, obtaining a new catalog: 4FGL-DR2 / 4LAC-DR2.
\textit{Initial set}:
3436 blazars including 80 TeV ones,
with 88 features.
\textit{Learning set}:
861 blazars including 71 TeV ones,
with 32 features.
688 / 173 blazars for each training / testing set.
\textit{Prediction set}:
1459  blazars including 73 TeV ones,
with 5 features.

\item [3FHL]
\textit{Initial set}:
1207 blazars including 74 TeV ones,
with 49 features.
Learning set:
506 blazars including 63 TeV blazars,
with 18 features.
404 / 102 blazars for each training / testing set.
\textit{Prediction set}:
540 blazars including 67 TeV ones,
with 2 features.

\item [3HSP]
\textit{Initial set}:
2013 blazars including 57 TeV ones,
with 13 features.
\textit{Learning set}:
1411 blazars including 53 TeV ones,
with 5 features.
1128 / 283 blazars for each training / testing set.
\textit{Prediction set}:
1771 blazars including 54 TeV ones,
with 2 features.

\item [2BIGB]
\textit{Initial set}:
1160 blazars including 57 TeV ones,
with 31 features.
\textit{Learning set}:
1040 blazars including 54 TeV ones,
with 7 features.
832 / 208 blazars for each training / testing set.
\textit{Prediction set}:
1040 blazars including 54 TeV ones,
with 3 features.
\end{description}

\subsection{logit, logistic, and performance of LR}
Processing the SML method we ascertain the result meeting condition of
$logistic \geqslant$ 80\% which gives 40 high-confidence candidates out of 150 4FGL-DR2 / 4LAC-DR2 candidates (see also Fig.~\ref{sed}, Tab.~\ref{IACTs}).
Among the 40 high-confidence candidates, 24 sources are in common with 3FHL candidates, 11 with 3HSP candidates, and 14 with 2BIGB candidates. We report the main results of SML as follows:

\begin{description}
\item [4FGL-DR2 / 4LAC-DR2]
\begin{equation}
\label{logit1}
logit^{\prime} = 4.808 \Gamma + 2.809 V_{F} + 3.889 \log f^{ph}_{\rm 7} - 3.34 z + 0.857 \log \nu_{\rm p}^{\rm s} + 20.244,
\end{equation}
where $logit^{\prime} = 0$ is ideal for $p_{\rm thre} = $40\%.
The optimal LR model is built on training 3,
with $AUC$ being 97\% on training 3 and 96\% on testing 3.
150 TeV blazar candidates are obtained from 1459 blazars in the prediction set, 
with $AUC$: 98\%, $FPR$: 11\%, and $TPR$: 99\%.
Only 1 TeV blazar is mistaken as a non-TeV one.

\item [3FHL]
\begin{equation}
\label{logit2}
logit^{\prime} = 0.116 \log f ^{\rm ph}_{\rm 2} - 0.628 z + 1.028,
\end{equation}
where $logit^{\prime} = 0$ is ideal for $p_{\rm thre} = $52\%,
the optimal LR model is built on training 5,
with $AUC$ being 89\% on training 5 and 87\% on testing 5.
126 TeV candidates are obtained from 540 blazars,
with $AUC$: 88\%, $FPR$: 27\%, and $TPR$: 88\%.
Only 8 TeV blazars are mistaken as non-TeV ones.

\item [3HSP]
\begin{equation}
\label{logit3}
logit^{\prime} = 0.306 FOM - 3.861 z - 0.173,
\end{equation}
where $logit^{\prime} = 0$ is ideal for $p_{\rm thre} = $61\%,
the optimal LR model is built on training 5,
with $AUC$ being 99\% on training 5 and 94\% on testing 5,
40 TeV candidates are obtained from 1771 blazars,
with $AUC$: 98\%, $FPR$: 2\%, and $TPR$: 91\%,
only 5 TeV blazars are mistaken as non-TeV ones.

\item [2BIGB]
\begin{equation}
\label{logit4}
logit^{\prime} = -0.016 E_{\rm P} + 0.248 FOM - 4.395z - 0.122
\end{equation}
where $logit^{\prime} = 0$ is ideal for $p_{\rm thre} = $48\%,
the optimal LR model is built on training 4,
with $AUC$ being 97\% on training 4 and 97\% on testing 4.
83 TeV candidates are obtained from 1040 blazars,
with $AUC$: 97\%, $FPR$: 8\%, and $TPR$: 96\%,
only 2 TeV blazars are mistaken as non-TeV ones.
\end{description}

The features of the 4 \textit{learning sets} are reported in Tab.~\ref{features},
where Col. (1) indicates the catalog information;
Col. (2) gives how many columns the features contain, and each column represents a feature;
Col. (3) and Col. (4) show the names and units of the features from the FITS version of the catalogs;
Col. (5) is the description of the feature.
The performance of the LR model on 4 learning sets is also shown in Tab.~\ref{LR_4FGL} to Tab.~\ref{LR_2BIGB},
where Col. (1) shows the datasets name;
Col. (2) the $AUC$, where the top is for training sets while the bottom for testing sets; 
Col. (3) $AUC$ difference between training sets and testing sets;
Col. (4) is the $p_{\rm thre}$;
Col. (5) is optimal features;
Col. (6) is optimal hyper-parameters,
note that the bold corresponds to minimum overfitting datasets,
which represents the data and parameters of {the optimal} LR model.

\begin{table*}[htbp]
\centering
\caption{Features from 4 catalogs for learning sets. \label{features}}
\resizebox{\textwidth}{!}{
\begin{tabular}{ccccc}
\cmidrule{1-5}    Catalog & Column & Feature & Unit  & Description \\
(1)  & (2)  & (3)  & (4) & (5)\\
\midrule
\multirow{20}[2]{*}{4FGL-DR2 / 4LAC-DR2} & 1     & Pivot\_Energy ($E_{\rm P}$) & GeV   & Energy at which error on differential flux is minimal \\
& 1     & Flux ($\log F^{\rm ph}$) & $\mathrm{cm^{-2} \cdot s^{-1}}$ & Integral photon flux from 1 to 100 GeV \\
& 1     & Energy\_Flux ($\log F$) & $\mathrm{erg \cdot cm^{-2} \cdot s^{-1}}$ & Energy flux from 100 MeV to 100 GeV\\
& 1     & PL\_Flux\_Density ($\log f_{\rm 1}$) & $\mathrm{cm^{-2} \cdot MeV^{-1} \cdot s^{-1}}$ & Differential flux at Pivot\_Energy in PowerLaw fit \\
& 1     & PL\_Index ($\alpha_{\rm 1}$) &      & Photon index when fitting with PowerLaw \\
& 1     & LP\_Flux\_Density ($\log f_{\rm 2}$) & $\mathrm{cm^{-2} \cdot MeV^{-1} \cdot s^{-1}}$ & Differential flux at Pivot\_Energy in LogParabola fit \\
& 1     & LP\_Index ($\alpha_{\rm 2}$) &       & Photon index at Pivot\_Energy when fitting with LogParabola \\
& 1     & LP\_beta ($\beta$)&       & Curvature parameter when fitting with LogParabola \\
& 1     & PLEC\_Flux\_Density ($\log f_{\rm 3}$)& $\mathrm{cm^{-2} \cdot MeV^{-1} \cdot s^{-1}}$ & Differential flux at Pivot\_Energy in PLSuperExpCutoff fit \\
& 1     & PLEC\_Index ($\Gamma$)&       & Low-energy photon index when fitting with PLSuperExpCutoff \\
& 1     & PLEC\_Expfactor ($a$)&       & Exponential factor when fitting with PLSuperExpCutoff \\
& 1     & PLEC\_Exp\_Index ($b$)&       & Exponential index when fitting with PLSuperExpCutoff \\
& 7     & Flux\_Band ($\log f_{\rm 1} \sim \log f_{\rm 7}$)& $\mathrm{cm^{-2} \cdot s^{-1}}$ & Integral photon flux in each spectral band \\
& 7     & nuFnu\_Band ($\log f ^{\rm ph}_{\rm 1} \sim \log f ^{\rm ph}_{\rm 7}$)& $\mathrm{erg \cdot cm^{-2} \cdot s^{-1}}$ & Spectral energy distribution over each spectral band \\
& 1     & Variability\_Index ($V$)&       & Likelihood difference between the flux fitted in each time interval and the average flux\\
& 1     & Frac\_Variability ($V_{\rm F}$)&       & Fractional variability computed from the fluxes in each year\newline{} \\
& 1     & Redshift ($z$)&       & Redshift \\
& 1     & nu\_syn ($\log \nu_{\rm p}^{\rm s}$)& Hz    & Synchrotron-peak frequency in observer frame \\
& 1     & nuFnu\_syn ($\log \nu_{\rm p}^{\rm s}f_{\nu_{\rm p}}^{\rm s}$)& $\mathrm{erg \cdot cm^{-2} \cdot s^{-1}}$ & Spectral energy distribution at synchrotron-peak frequency \\
& 1     & Highest\_energy ($E_{\rm H}$)& GeV   & Highest energy among events probably coming from the source \\
\midrule
\multirow{12}[1]{*}{3FHL} & 1     & Pivot Energy ($E_{\rm P}$)& GeV   & Energy at which error on differential flux is minimal \\
& 1     & Flux Density ($\log f$) & $\mathrm{cm^{-2} \cdot GeV^{-1} \cdot s^{-1}}$ & Differential flux at Pivot Energy \\
& 1     & Flux ($\log F ^{\rm ph}$) & $\mathrm{cm^{-2} \cdot s^{-1}}$ & Integral photon flux from 10 GeV to 1 TeV $\mathrm{erg\ cm^{-2} \cdot s^{-1}}$ \\
& 1     & Energy Flux ($\log F$)& $\mathrm{erg \cdot cm^{-2} \cdot s^{-1}}$ & Energy flux from 10 GeV to 1 TeV \\
& 1     & PowerLaw Index ($\alpha_{\rm 1}$) &       & Photon index when fitting with power law \\
& 1     & Spectral Index ($\alpha_{\rm 2}$) &       & Photon index at Pivot Energy when fitting with LogParabola \\
& 1     & beta ($\beta$) &       & Curvature parameter when fitting with LogParabola \\
& 4     & Flux Band ($\log f_{\rm 1} \sim \log f_{\rm 4}$)& $\mathrm{cm^{-2} \cdot s^{-1}}$ & Integral photon flux in each spectral band \\
& 4     & nuFnu\_Band ($\log f ^{\rm ph}_{\rm 1} \sim \log f ^{\rm ph}_{\rm 4}$)& $\mathrm{erg \cdot cm^{-2} \cdot s^{-1}}$ & Spectral energy distribution over each spectral band \\
& 1     & HEP energy ($E_{\rm H}$) & GeV   & Highest energy among events probably coming from the source \\
& 1     & Redshift ($z$)&       & Redshift \\
& 1     & NuPeak obs ($\log \nu_{\rm p}^{\rm s}$)& Hz    & Synchrotron-peak frequency in observer frame \\
\midrule
\multirow{5}[1]{*}{3HSP} & 1     & radio flux density ($\log f_{\rm R}$)& mJy   & Radio flux density from the NVSS or FIRST catalog \\
& 1     & X-ray flux flux density ($\log f_{\rm X}$)& $\mu$Jy   & X-ray flux density at 1keV \\
& 1     & nu\_syn ($\log \nu_{\rm p}^{\rm s}$)& Hz    & Synchrotron-peak frequency in observer frame \\
& 1     & Redshift ($z$)&       & Redshift \\
& 1     & FOM ($FOM$)  &       & The figure of merit parameter, which is related to the likelihood of GeV / TeV detectability \\
\midrule
\multirow{7}[2]{*}{2BIGB} & 1     & N0 ($\log f$)& $\mathrm{cm^{-2} \cdot MeV^{-1} \cdot s^{-1}}$ & Differential flux at Pivot Energy in PowerLaw fit\\
& 1     & Gamma ($\alpha$)&       & Photon index when fitting with PowerLaw \\
& 1     & $\rm F_{0.5-500 GeV}$ ($F ^{\rm ph}$) &       & Integrated photon flux from 500 MeV to 500 GeV\\
& 1     & E0 ($E_{\rm P}$) & GeV   & Energy at which error on differential flux is minimal \\
& 1     & nu\_syn ($\log \nu_{\rm p}^{\rm s}$)& Hz    & Synchrotron-peak frequency in observer frame from 3HSP \\
& 1     & Redshift ($z$)&       & Redshift from 3HSP \\
& 1     & FOM ($FOM$)  &       & The same as in 3HSP \\
\bottomrule
\end{tabular}
}
\end{table*}%

\begin{table*}[htbp]
\scriptsize
\centering
\caption{LR model and performance for the learning dataset of 4FGL-DR2 / 4LAC-DR2}
\begin{tabular}{cccccc}
\midrule
\midrule
Partition & $AUC$  & Overfitting & $p_{\rm thre}$ & Features  & Hyper-parameters \\
(1)  & (2)  & (3)  & (4) & (5)  & (6) \\
\midrule
\multirow{2}[0]{*}{1} & 96.9\%  & \multirow{2}[0]{*}{0.9\%} & \multirow{2}[0]{*}{53.6\%} & \multirow{2}[0]{*}{\makecell[c]{$E_{\rm p}, {\log}F^{\rm ph}_{\rm 1}, {\log}F_{\rm 1}, \mathit\Gamma, V, V_{\rm f}, {\log} f^{\rm ph}_{\rm 1},$\\$ {\log} f^{\rm ph}_{\rm 2}, {\log} f^{\rm ph}_{\rm 4},{\log} f^{\rm ph}_{\rm 6}, {\log} f^{\rm ph}_{\rm 7}, z, {\log} \nu_{\rm p}^{\rm s}$}} & \multirow{2}[0]{*}{\makecell[c]{C: \emph{1}, max\_iter: \emph{500}, penalty: \emph{l2}, \\ solver: \emph{sag}, tol:  $\mathit10^{-4}$}} \\
& 95.6\%  &       &       &       &  \\
\midrule
\multirow{2}[0]{*}{2} & 97.2\%  & \multirow{2}[0]{*}{1.2\%} & \multirow{2}[0]{*}{53.5\%} & \multirow{2}[0]{*}{\makecell[c]{$E_{\rm p}, F_{\rm 1}, \mathit{\alpha_{\rm 1}, \Gamma}, V,$ \\ $ V_{\rm f}, f^{\rm ph}_{\rm 7}, z, \log \nu_{\rm p}^{\rm s}$}} & \multirow{2}[0]{*}{\makecell[c]{C: \emph{1}, max\_iter: \emph{100}, penalty: \emph{l1}, \\ solver: \emph{liblinear}, tol: $\mathit {10}^{\rm -6}$}} \\
& 96.0\%  &       &       &       &  \\
\midrule
\multirow{2}[0]{*}{\textbf{3}} & \textbf{96.8\%} & \multirow{2}[0]{*}{\textbf{0.8\%}} & \multirow{2}[0]{*}{\textbf{53.4\%}} & \multirow{2}[0]{*}{\makecell[c]{$\bm{\mathit{\Gamma, V_{\rm f}, \log f^{\rm ph}_{\rm 7}, z, \log \nu_{\rm p}^{\rm s}}}$}}& \multirow{2}[0]{*}{\makecell[c]{\textbf{C: \emph{1}, max\_iter: \emph{500}, penalty: \emph{l1},} \\ \textbf{solver: \emph{saga}, tol: ${\bm{\mathit {10}^{\rm -6}}}$}}} \\
& \textbf{96.0\%} &       &       &       &  \\
\midrule
\multirow{2}[0]{*}{4} & 96.0\%  & \multirow{2}[0]{*}{-1.8\%} & \multirow{2}[0]{*}{53.0\%} & \multirow{2}[0]{*}{\makecell[c]{$\log F_{\rm 1}, \mathit{\alpha_{\rm 1}, \Gamma}, a, V_{\rm f},$ \\ ${\log} f_{\rm 7}, \log F^{\rm ph}_{\rm 7}, z, \log \nu_{\rm p}^{\rm s}$}} & \multirow{2}[0]{*}{\makecell[c]{C: \emph{0.1}, max\_iter: \emph{100}, penalty: \emph{l2}, \\ solver: \emph{newton-cg}, tol: $\mathit {10}^{-6}$}} \\
& 98.0\%  &       &       &       &  \\
\midrule
\multirow{2}[0]{*}{5} & 97.5\%  & \multirow{2}[0]{*}{5.0\%} & \multirow{2}[0]{*}{52.2\%} & \multirow{2}[0]{*}{\makecell[c]{$E_{\rm p}, \mathit {\alpha_{\rm 1}, \alpha_{\rm 3}}, V_{\rm f}, \log f_{\rm 3}, \log f_{\rm 5}, $ \\ $\log f_{\rm 7}, \log f^{ph}_{\rm 7}, z, \log \nu_{\rm p}^{\rm s}$}} & \multirow{2}[0]{*}{\makecell[c]{C: \emph{1}, max\_iter: \emph{100}, penalty: \emph{l2}, \\ solver: \emph{newton-cg}, tol: $\mathit {10}^{-6}$}} \\
& 92.5\%&       &       &       &  \\
\midrule
\label{LR_4FGL}
\end{tabular}
\end{table*}

\begin{table*}[htbp]
\scriptsize
\centering
\caption{LR model and performance for the learning set of 3FHL}
\begin{tabular}{cccccc}
\midrule
\midrule
Partition & $AUC$   & Overfitting & $p_{\rm thre}$ & Features   & Hyper-parameters \\
(1)  & (2)  & (3)  & (4) & (5)  & (6) \\
\midrule
\multirow{2}[0]{*}{1} & 87.8\%  & \multirow{2}[0]{*}{-1.9\%} & \multirow{2}[0]{*}{53.6\%} & \multirow{2}[0]{*}{$\log f ^{ph}_{2}, z$} & 
\multirow{2}[0]{*}{\makecell[c]{C: \emph{0.01}, l1\_ratio: \emph{0.4}, max\_iter: \emph{100}, \\ penalty: \emph{elasticnet}, solver: \emph{saga}, tol: $\mathit{10}^{-6}$}} \\
& 89.7\%  &       &       &  &  \\
\midrule
\multirow{2}[0]{*}{2} & 88.0\%  & \multirow{2}[0]{*}{-7.4\%} & \multirow{2}[0]{*}{53.5\%} & \multirow{2}[0]{*}{$\log f ^{ph}_{\rm 2}, z$} & \multirow{2}[0]{*}{\makecell[c]{C: \emph{0.01}, l1\_ratio: \emph{0.3}, max\_iter: \emph{100}, \\ penalty: \emph{elasticnet}, solver: \emph{saga}, tol: $\mathit{10}^{-6}$}} \\
& 95.5\%  &       &       &  &  \\
\midrule
\multirow{2}[0]{*}{3} &88.7\%  & \multirow{2}[0]{*}{5.5\%} & \multirow{2}[0]{*}{53.4\%} & \multirow{2}[0]{*}{$\log f ^{ph}_{\rm 2}, z$} & \multirow{2}[0]{*}{\makecell[c]{C: \emph{0.01}, l1\_ratio: \emph{0.4}, max\_iter: \emph{100}, \\ penalty: \emph{elasticnet}, solver: \emph{saga}, tol: $\mathit{10}^{-6}$}} \\
& 83.1\%  &       &       &  &  \\
\midrule
\multirow{2}[0]{*}{4} & 89.6\%  & \multirow{2}[0]{*}{3.6\%} & \multirow{2}[0]{*}{52.9\%} & \multirow{2}[0]{*}{$\log f ^{ph}_{\rm 2}, z$} & \multirow{2}[0]{*}{\makecell[c]{C: \emph{0.01}, l1\_ratio: \emph{0.2}, max\_iter: \emph{100}, \\ penalty: \emph{elasticnet}, solver: \emph{saga}, tol: $\mathit{10}^{-6}$}} \\
& 86.0\%  &       &       &  &  \\
\midrule
\multirow{2}[0]{*}{5} & \textbf{89.2\%} & \multirow{2}[0]{*}{\textbf{2.3\%}} & \multirow{2}[0]{*}{\textbf{52.2\%}} & \multirow{2}[0]{*}{$\log f ^{ph}_{\rm 2}, z$} & \multirow{2}[0]{*}{\makecell[c]{\textbf{C: \emph{0.01}, l1\_ratio: \emph{0.4}, max\_iter: \emph{100}}, \\ \textbf{penalty: \emph{elasticnet}, solver: \emph{saga}, tol: $\bm {\mathit{10}^{-6}}$}}} \\
& \textbf{87.0\%} &       &       & &  \\
\midrule
\label{LR_3FHL}
\end{tabular}
\end{table*}%

\begin{table*}[htbp]
\scriptsize
\centering
\caption{LR model and performance for the learning set of 3HSP}%
\begin{tabular}{cccccc}
\midrule
\midrule
Partition & \multicolumn{1}{c}{$AUC$} & Overfitting & $p_{\rm thre}$ & Features   & Hyper-parameters \\
(1)  & (2)  & (3)  & (4) & (5)  & (6) \\
\midrule
\multirow{2}[0]{*}{1} & 98.3\%  & \multirow{2}[0]{*}{-1.2\%} & \multirow{2}[0]{*}{46.9\%} & \multirow{2}[0]{*}{$\log f_{\rm X}, \log \nu ^{\rm s}_{\rm p}, FOM, z$} & \multirow{2}[0]{*}{\makecell[c]{C: \emph{0.01}, l1\_ratio: \emph{0.7}, max\_iter: \emph{100}, \\ penalty: \emph{elasticnet}, solver: \emph{saga}, tol: $\textit{10} ^{-6}$}} \\
& 99.5\%  &       &       &       &  \\
\midrule
\multirow{2}[0]{*}{2} & 98.4\%  & \multirow{2}[0]{*}{-0.5\%} & \multirow{2}[0]{*}{60.2\%} & \multirow{2}[0]{*}{$FOM, z$} & \multirow{2}[0]{*}{\makecell[c]{C: \emph{0.01}, max\_iter: \emph{100}, \\penalty: \emph{l2}, solver: \emph{liblinear}, tol: $\textit{10} ^{-6}$}} \\
& 98.9\%  &       &       &       &  \\
\midrule
\multirow{2}[0]{*}{3} & 98.6\%  & \multirow{2}[0]{*}{-0.5\%} & \multirow{2}[0]{*}{47.6\%} & \multirow{2}[0]{*}{$FOM, z$} & \multirow{2}[0]{*}{\makecell[c]{C: \emph{0.001}, max\_iter: \emph{100}, \\penalty: \emph{l2}, solver: \emph{newton-cg}, tol: $\textit{10} ^{-6}$}} \\
& 99.1\%  &       &       &       &  \\
\midrule
\multirow{2}[0]{*}{4} & 98.4\%  & \multirow{2}[0]{*}{-1.4\%} & \multirow{2}[0]{*}{58.2\%} & \multirow{2}[0]{*}{$FOM, z$} & \multirow{2}[0]{*}{\makecell[c]{C: \emph{0.1}, max\_iter: \emph{100}, \\penalty: \emph{l2}, solver: \emph{newton-cg}, tol: $\textit{10} ^{-6}$}} \\
& 99.8\%  &       &       &       &  \\
\midrule
\multirow{2}[0]{*}{\textbf{5}} & \textbf{99.2\%} & \multirow{2}[0]{*}{\textbf{5.7\%}} & \multirow{2}[0]{*}{\textbf{60.8\%}} & \multirow{2}[0]{*}{$\bm {\mathit{FOM, z}}$} & \multirow{2}[0]{*}{\makecell[c]{\textbf{C: \emph{0.01}, max\_iter: \emph{100}}, \\\textbf{penalty: \emph{l2}, solver: \emph{liblinear}, tol: $\textit{10} ^{ -6}$}}} \\
& \textbf{93.5\%} &       &       &       &  \\
\midrule
\end{tabular}
\label{LR_3HSP}
\end{table*}%

\begin{table*}[htbp]
\scriptsize
\centering
\caption{LR model and performance for the learning set of 2BIGB}
\label{LR_2BIGB}
\scalebox{1}{
\begin{tabular}{cccccc}
\midrule
\midrule
Partition & \multicolumn{1}{c}{$AUC$} & Overfitting & $p_{\rm thre}$ & Features   & Hyper-parameters \\
(1)  & (2)  & (3)  & (4) & (5)  & (6) \\
\midrule
\multirow{2}[0]{*}{1} & 98.6\%  & \multirow{2}[0]{*}{0.2\%} & \multirow{2}[0]{*}{44.7\%} & \multirow{2}[0]{*}{$\mathit\alpha, F, \log \nu ^{s}_{p}, FOM, z$} & \multirow{2}[0]{*}{\makecell[c]{C: \emph{0.01}, max\_iter: \emph{100}, penalty: \emph{l2}, \\ solver: \emph{newton-cg}, tol: $\textit{10} ^{-6}$}} \\
& 98.4\%  &       &       &       &  \\
\midrule
\multirow{2}[0]{*}{2} & 97.7\%  & \multirow{2}[0]{*}{6.0\%} & \multirow{2}[0]{*}{52.4\%} & \multirow{2}[0]{*}{$E_{\rm p}, FOM, z$} & \multirow{2}[0]{*}{\makecell[c]{C: \emph{0.01}, max\_iter: \emph{100}, penalty: \emph{l2}, \\solver: \emph{newton-cg}, tol: $\textit{10} ^{-6}$}} \\
& 91.7\%  &       &       &       &  \\
\midrule
\multirow{2}[0]{*}{3} &98.2\% & \multirow{2}[0]{*}{6.3\%} & \multirow{2}[0]{*}{49.1\%} & \multirow{2}[0]{*}{$FOM, z$} & \multirow{2}[0]{*}{\makecell[c]{C: \emph{0.01}, max\_iter: \emph{100}, penalty: \emph{l2}, \\ solver: \emph{newton-cg}, tol: $\textit{10} ^{-6}$}} \\
& 91.8\%  &       &       &       &  \\
\midrule
\multirow{2}[0]{*}{4} & \textbf{97.2\%} & \multirow{2}[0]{*}{\textbf{0.1\%}} & \multirow{2}[0]{*}{48.1\%} & \multirow{2}[0]{*}{$\bm {E_{\rm p}, FOM, z}$} & \multirow{2}[0]{*}{\makecell[c]{C: \emph{0.01}, l1\_ratio: \emph{0.1}, max\_iter: \emph{100},\\  penalty: \emph{elasticnet}, solver: \emph{saga}, tol: $\textit{10} ^{-6}$}} \\
& \textbf{97.1\%} &       &       &       &  \\
\midrule
\multirow{2}[0]{*}{5} & 96.7\%  & \multirow{2}[0]{*}{-3.0\% } & \multirow{2}[0]{*}{50.3\%} & \multirow{2}[0]{*}{$FOM, z$} & \multirow{2}[0]{*}{\makecell[c]{C: \emph{0.01}, max\_iter: \emph{100}, penalty: \emph{l2}, \\ solver: \emph{newton-cg}, tol: $\textit{10}^{-6}$}} \\
& 99.7\%  &       &       &       &  \\
\bottomrule
\end{tabular}}
\end{table*}%

\begin{figure*}[htbp]
\centering
\includegraphics[width = 5 in]{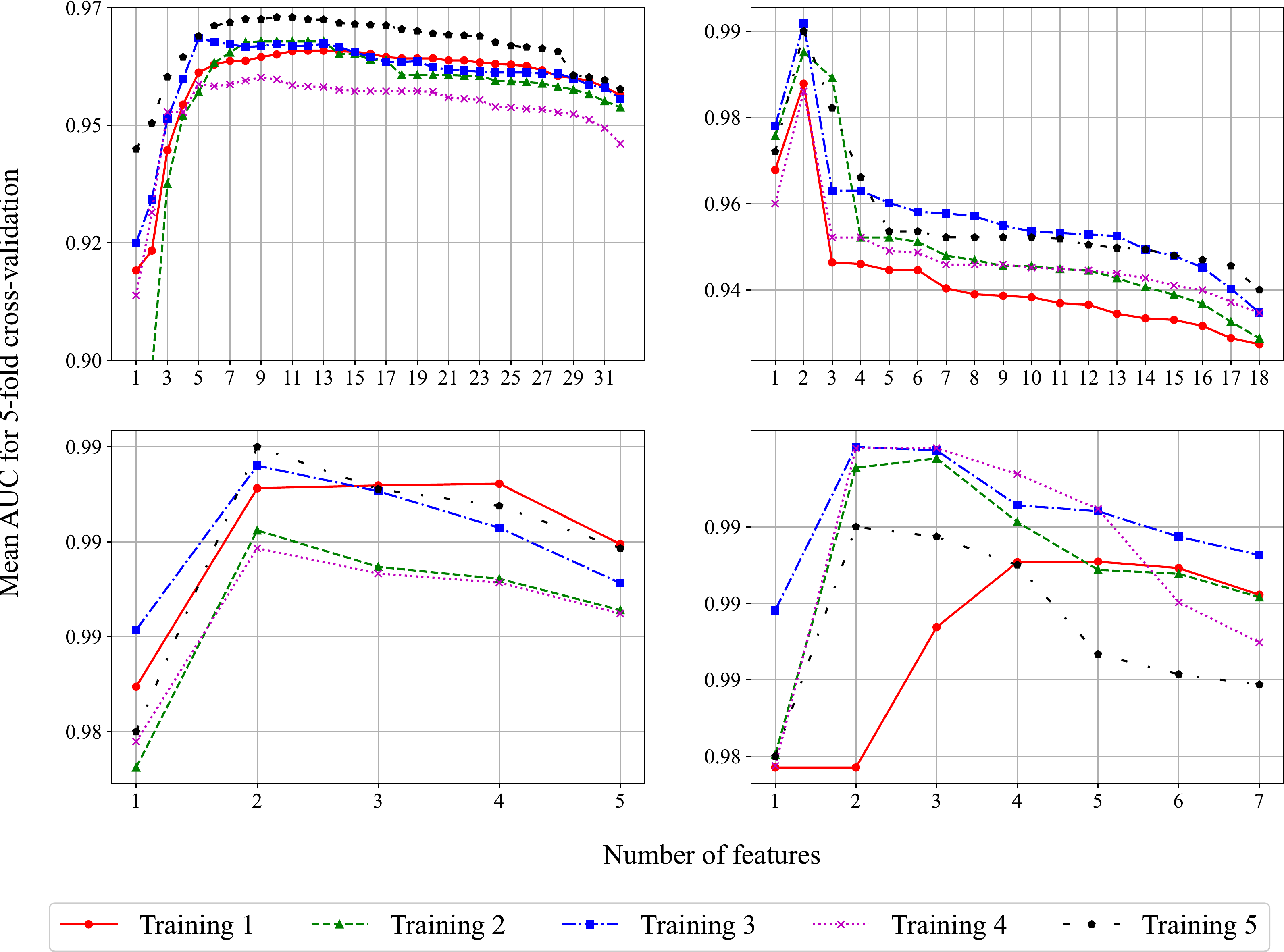}
\caption{
SFS for 4 learning sets,
where the X-axis is the number of parameters, and the Y-axis is the concentration of AUC on the 5 validation sets. 5 dotted lines in different colors indicate
training 1 $\sim$ 5.
Top left panel: 4FGL-DR2 / 4LAC-DR2;
Top right panel: 3FHL;
Bottom left panel: 3HSP;
Bottom right panel: 2BIGB.}
\label{sfs}
\end{figure*}

\begin{figure*}[htbp]
\centering
\includegraphics[width = 5 in]{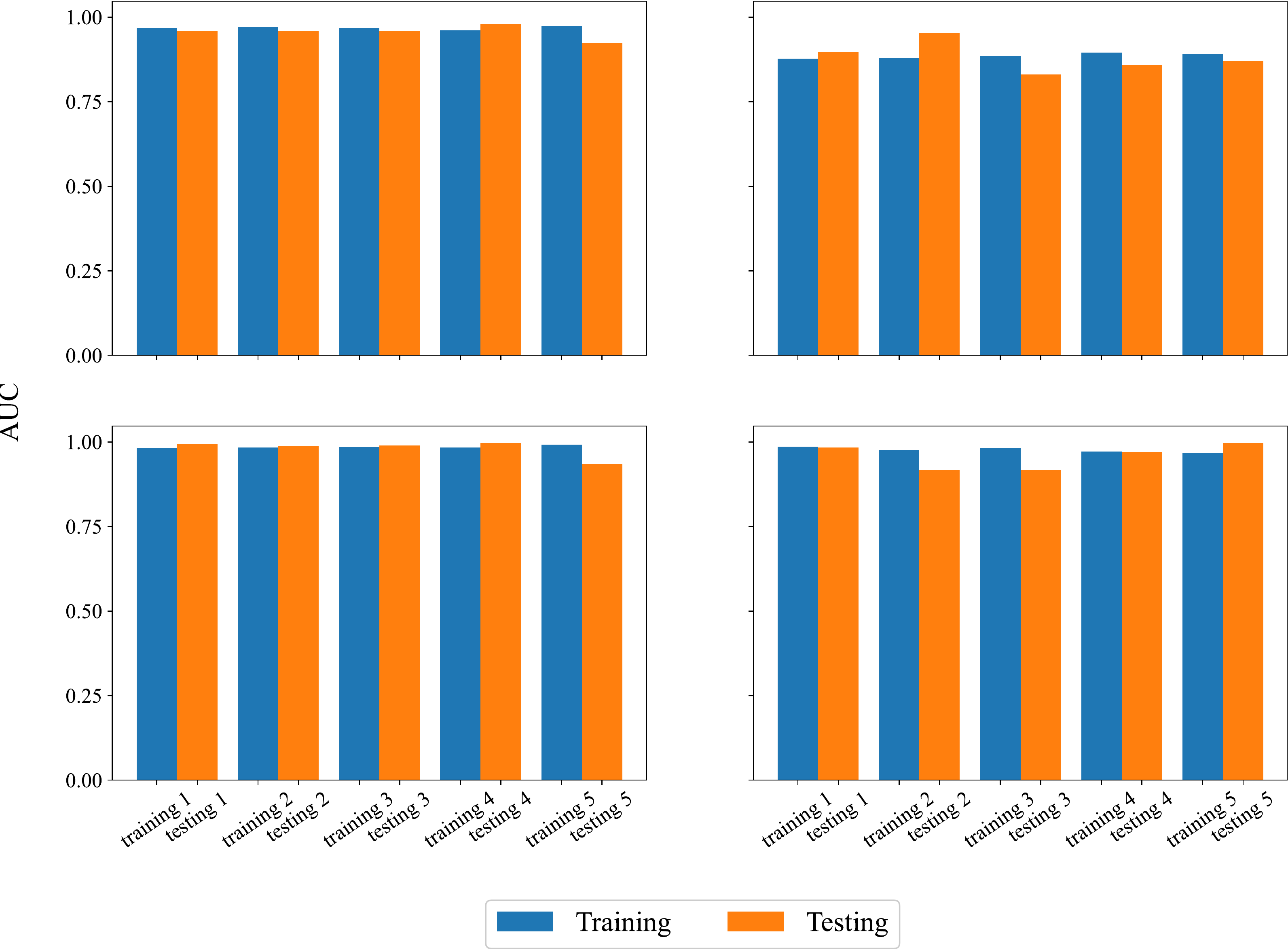}
\caption{
$AUC$ for 4 learning sets,
where blue histograms denote the training sets, and orange histograms represent the testing sets.
Top left panel: 4FGL-DR2 / 4LAC-DR2;
Top right panel: 3FHL;
Bottom left panel: 3HSP;
Bottom right panel: 2BIGB.
}
\label{ComparingAUC}
\end{figure*}

\begin{figure*}[htbp]
\centering
\includegraphics[width = 5 in]{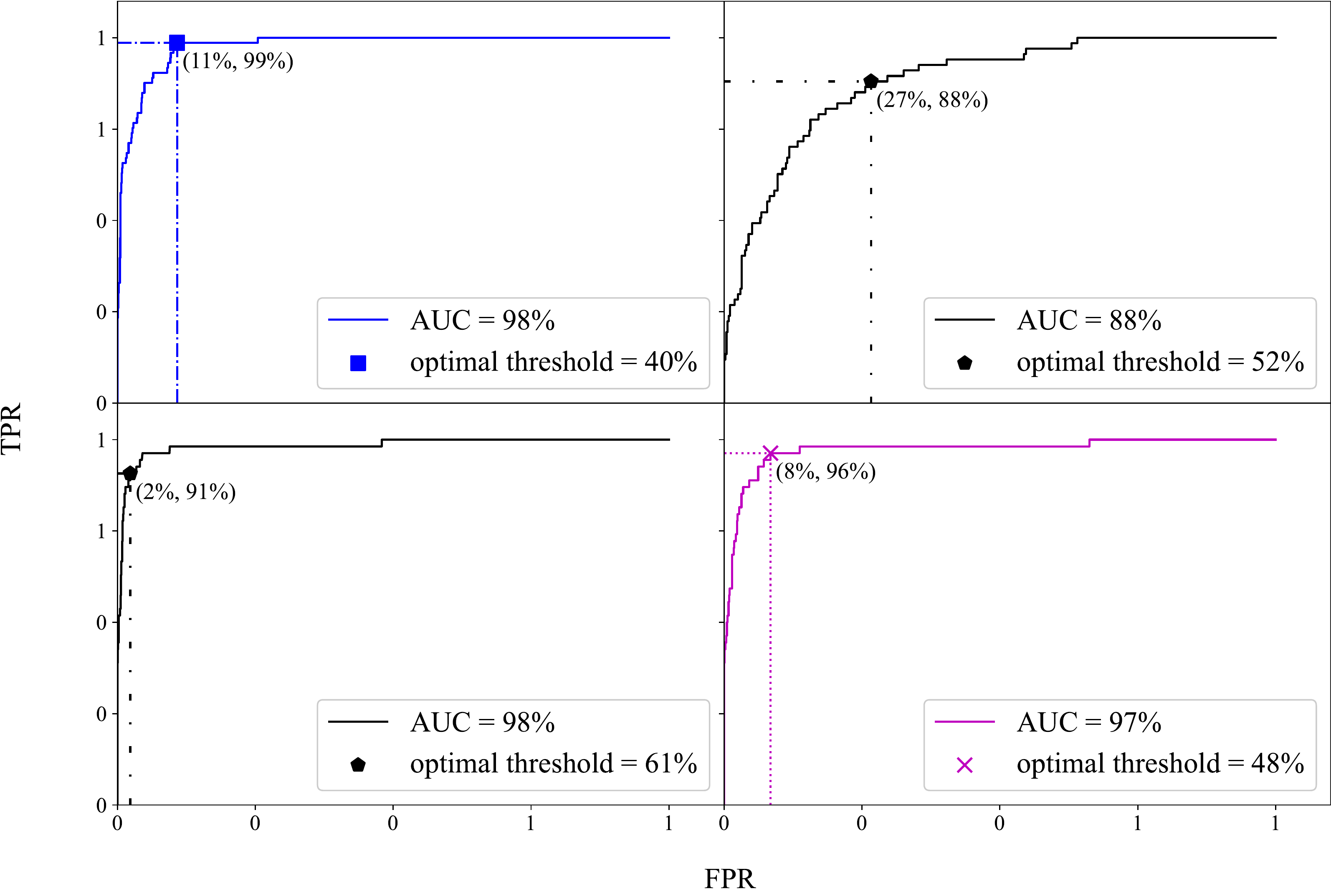}
\caption{ROC curve for 4 predicting datasets.
Top left panel: 4FGL-DR2 / 4LAC-DR2;
Top right panel: 3FHL;
Bottom left panel: 3HSP;
Bottom right panel: 2BIGB.}
\label{roc}
\end{figure*}

\begin{figure*}[htbp]
\centering
\includegraphics[width =5 in]{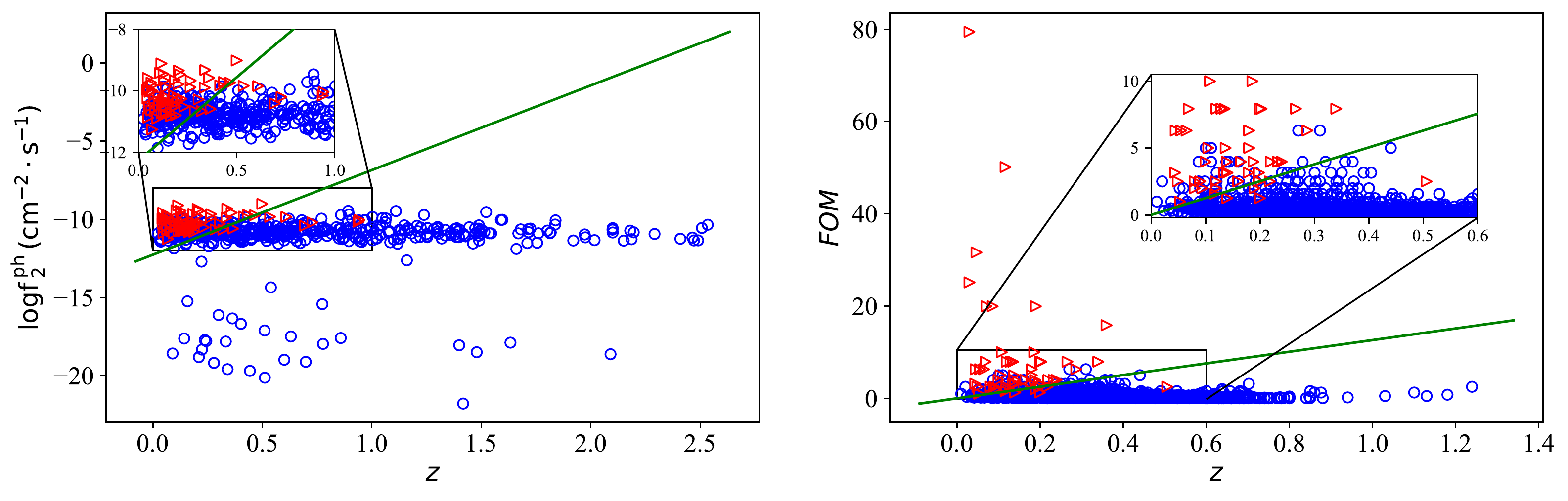}
\includegraphics[width = 2.8 in]{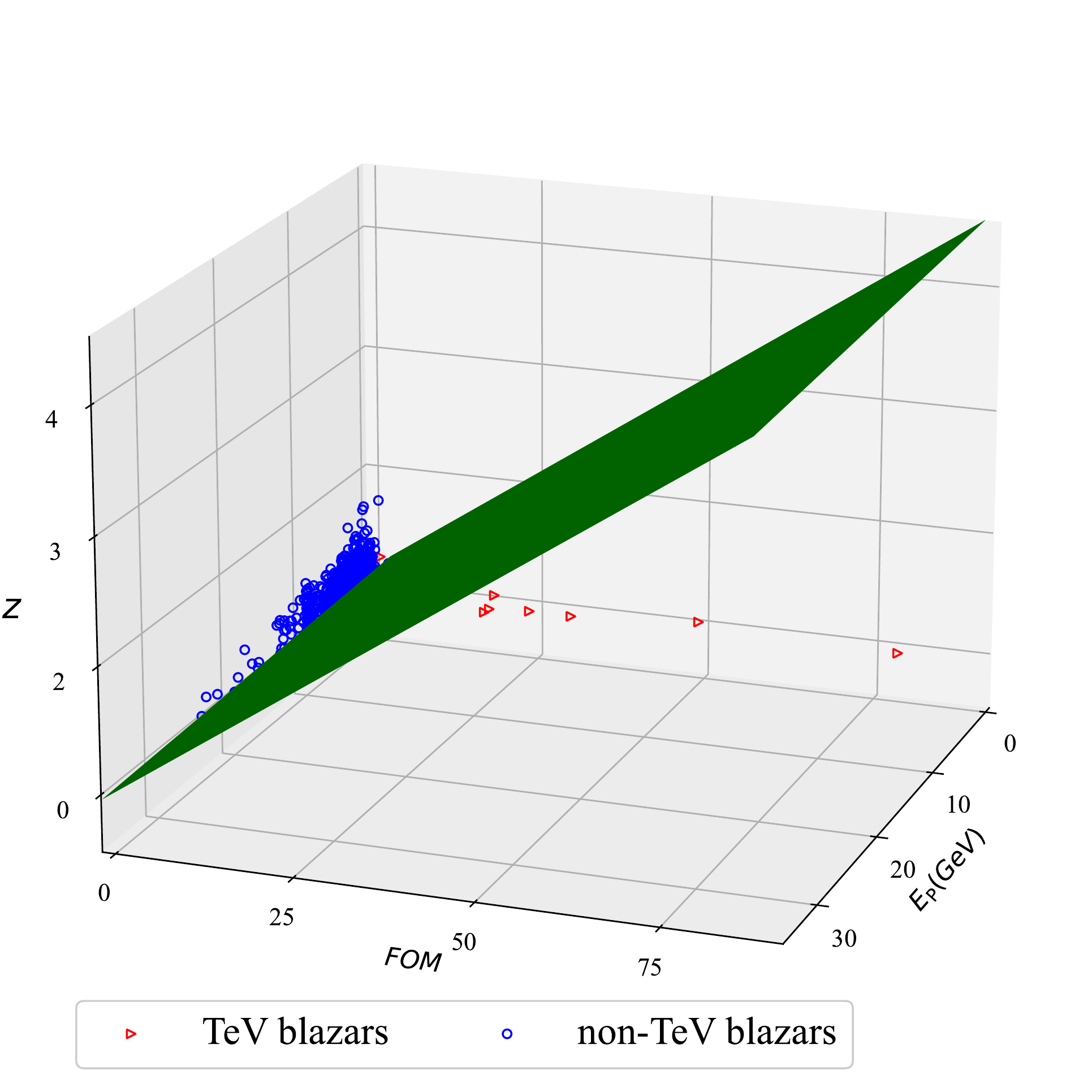}
\includegraphics[width = 2.8 in]{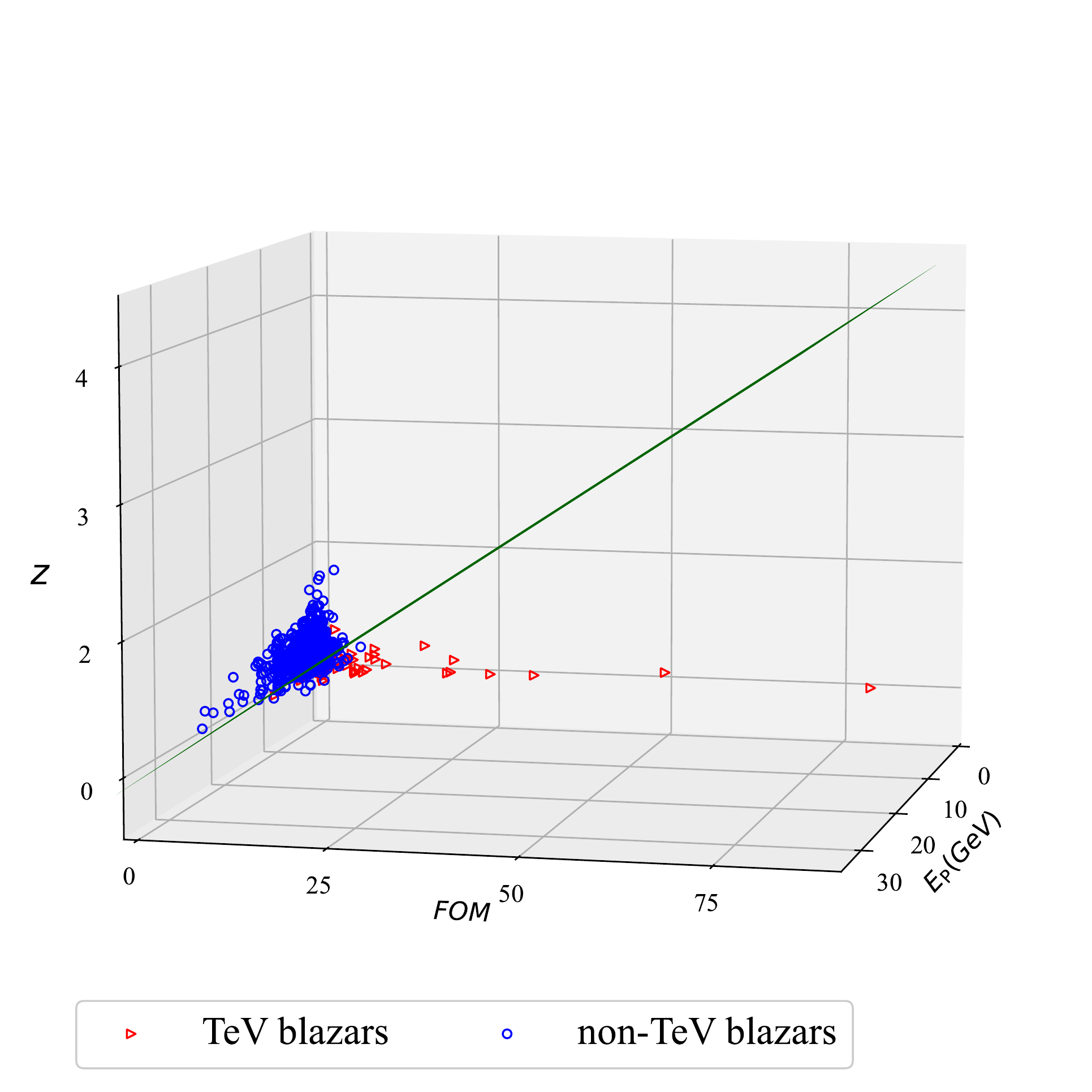}
\caption{Classify boundary and part of predicting datasets of 3FHL, 3HSP, and 2BIGB,
where the red triangle stands for TeV blazars, the blue circle stands for non-TeV ones, and the green line/plane stands for classification boundary.
Top left panel: 3FHL;
Top right panel: 3HSP;
Bottom left panel: 2BIGB on the view of non-TeVs side;
Bottom right panel: 2BIGB on the view along the edge of the green plane.}
\label{classify}
\end{figure*}

\begin{figure*}[htbp]
\centering
\includegraphics[width = 2.5 in]{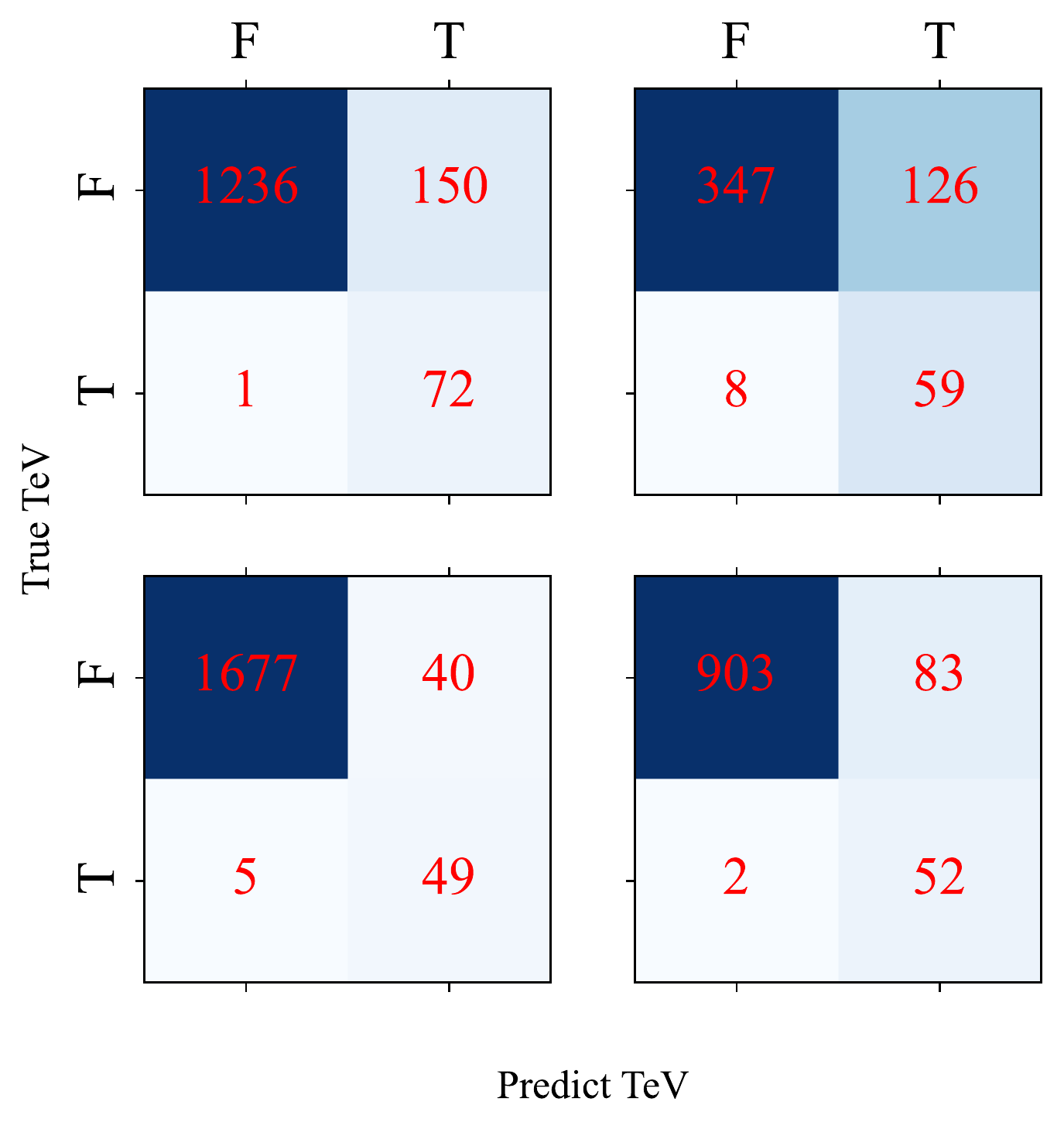}
\caption{Confusion matrix for 4 learning sets,
where the coordinate value `T' stands for `True', and `F' stands for `False'.
Top left panel: 4FGL-DR2 / 4LAC-DR2;
Top right panel: 3FHL;
Bottom left panel: 3HSP;
Bottom right panel: 2BIGB.
For each panel:
True positives (top left),
false negatives (top right),
false positives (bottom left),
and true negatives (bottom right).}
\label{cmat}
\end{figure*}

We also visualize part of the results in figures:
Fig.~\ref{sfs} shows feature selection;
Fig.~\ref{ComparingAUC} is $AUC$ of the training / testing sets;
Fig.~\ref{roc} illustrates \textit{ROC} curve, $AUC$, and $p_{\rm thre}$ of the prediction sets;
Fig.~\ref{classify} visualize the linear boundary of 3FHL, 3HSP, and 2BIGB.
Fig.~\ref{cmat} is the confuse matrix.

\section{Potential targets for ground-based Cherenkov detectors}
Among the 40 high-confidence TeV candidates, we consider the potential detectable targets of IACTs and EAS arrays for the whole year of 2023.
We employ two criteria to determine whether a source is detectable: (\romannumeral1) Its flux density is higher than the detector sensitivity curve at the energy band $\geqslant$ 1TeV; (\romannumeral2) It could pass by the detectable sky region of the detectors.

\subsection{SEDs considering EBL correction}
\begin{figure*}[htbp]
\centering
\includegraphics[width = 3.5 in]{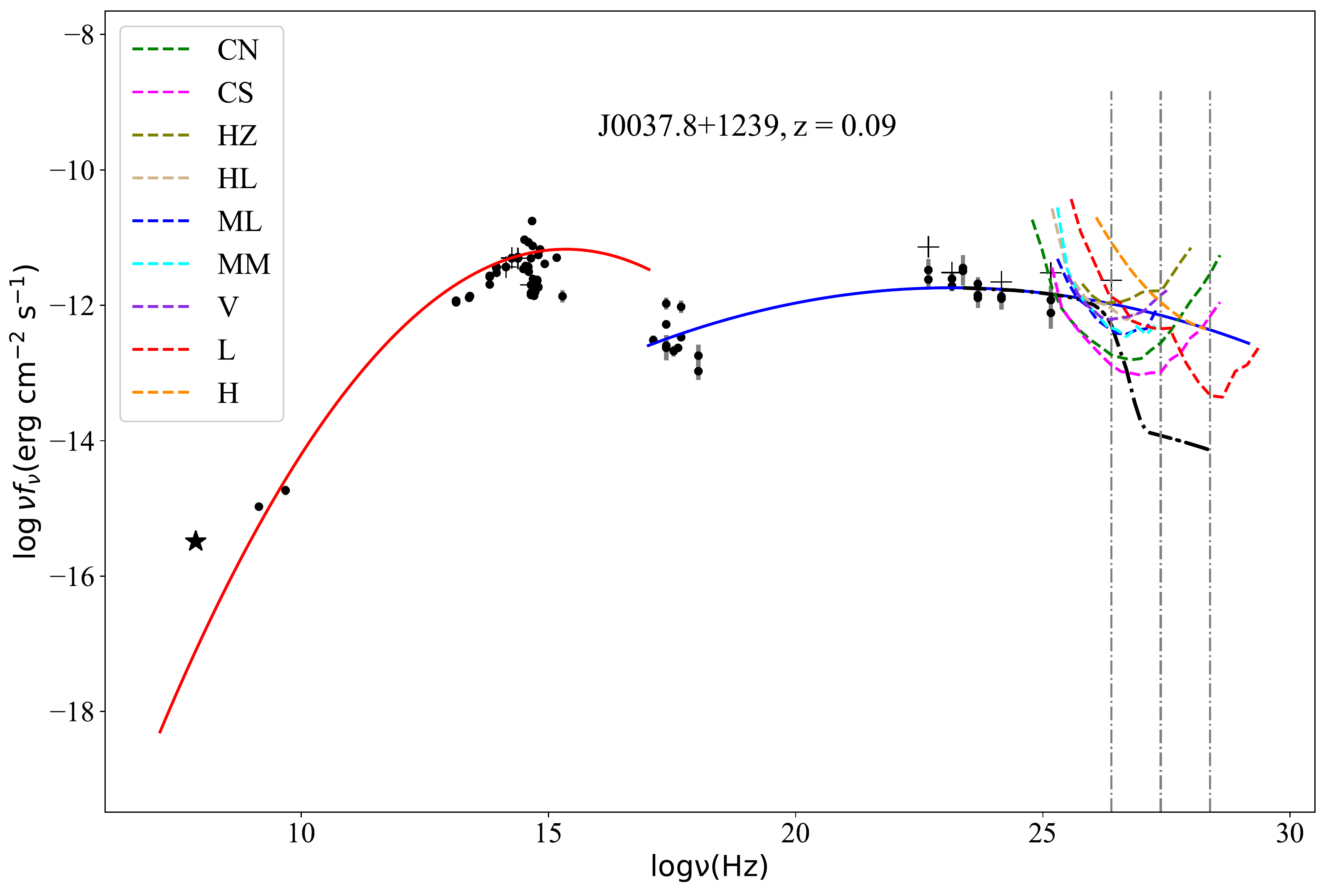}
\vspace{0.2in}
\includegraphics[width = 3.5 in]{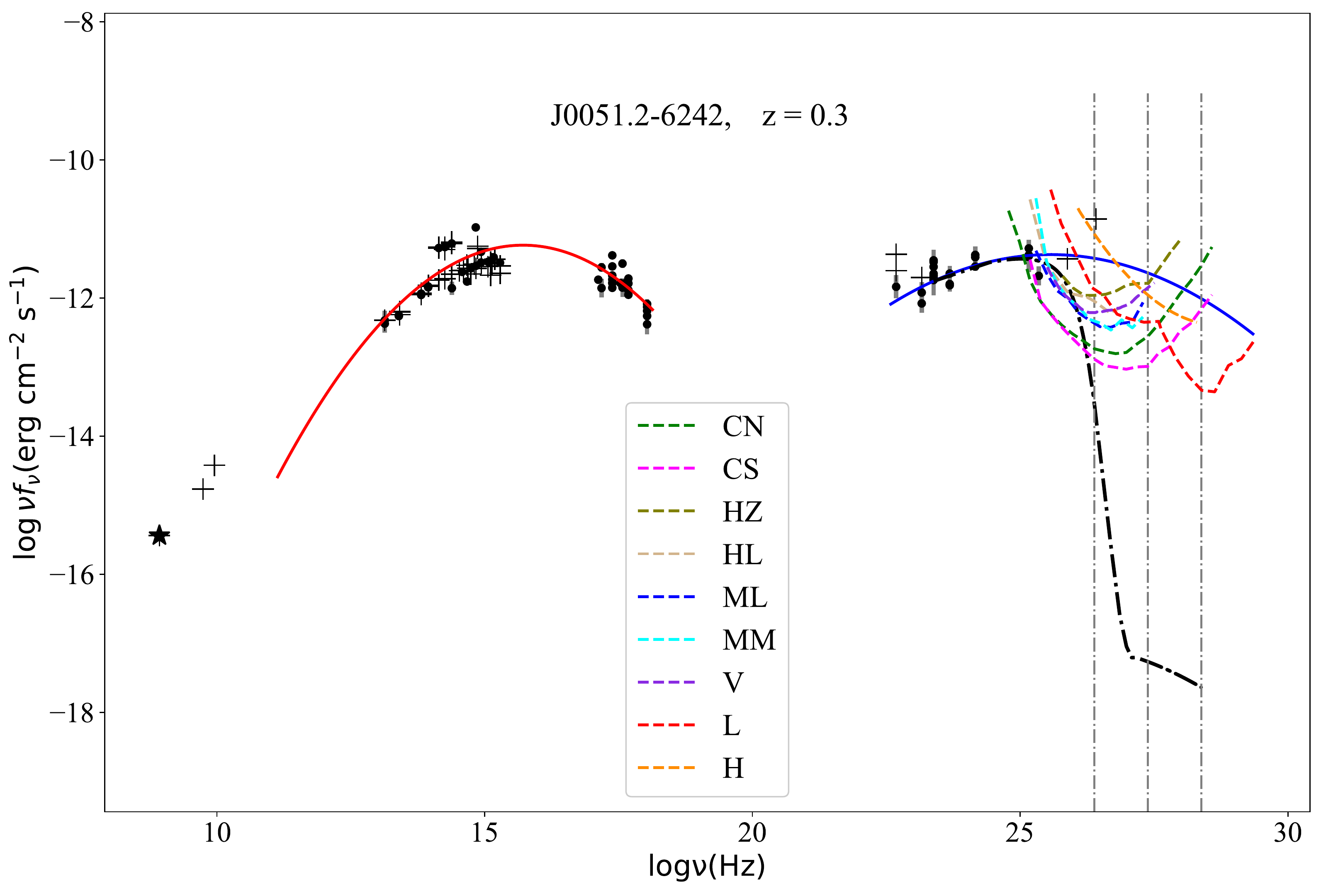}
\vspace{0.2in}
\includegraphics[width = 3.5 in]{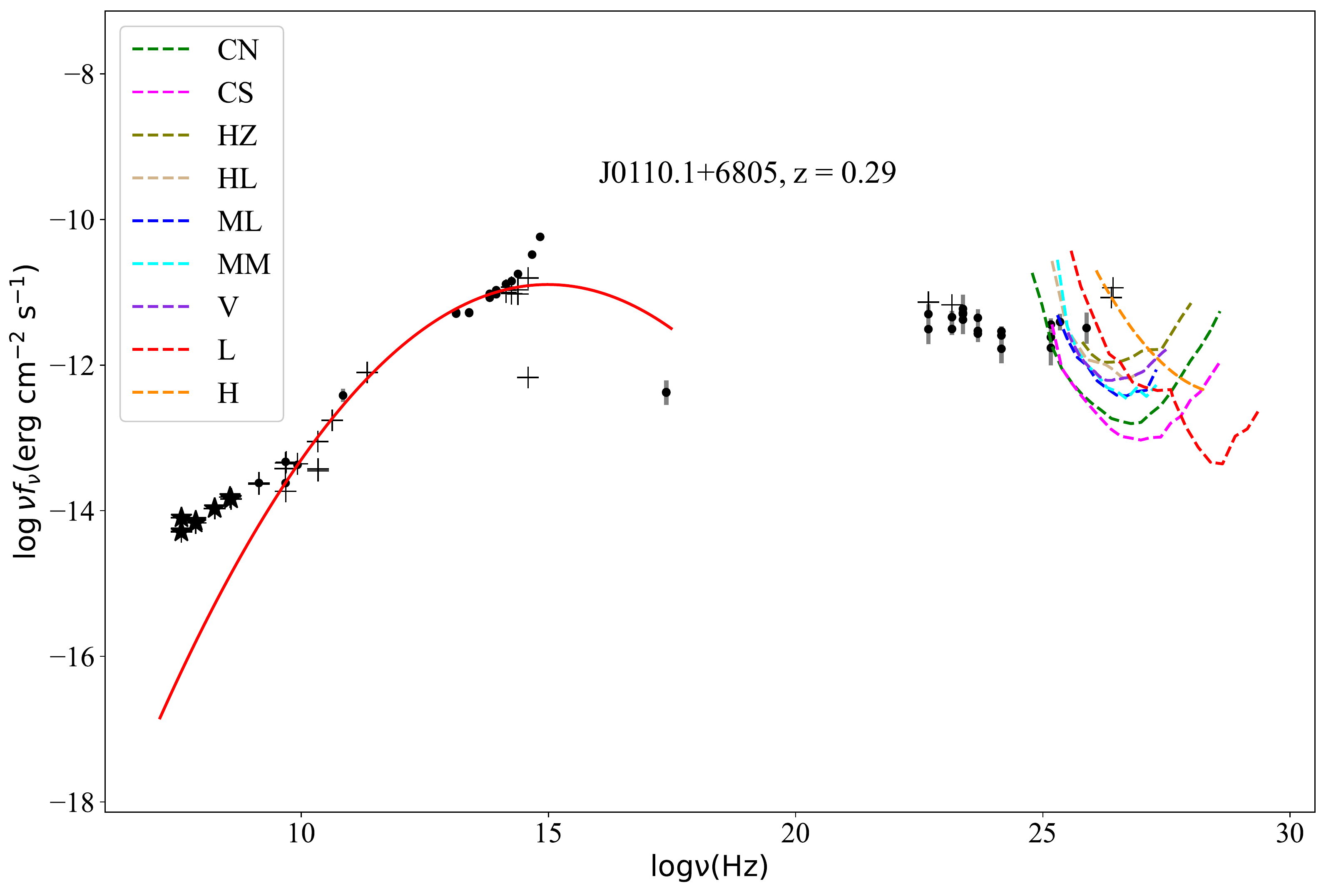}
\vspace{0.2in}
\includegraphics[width = 3.5 in]{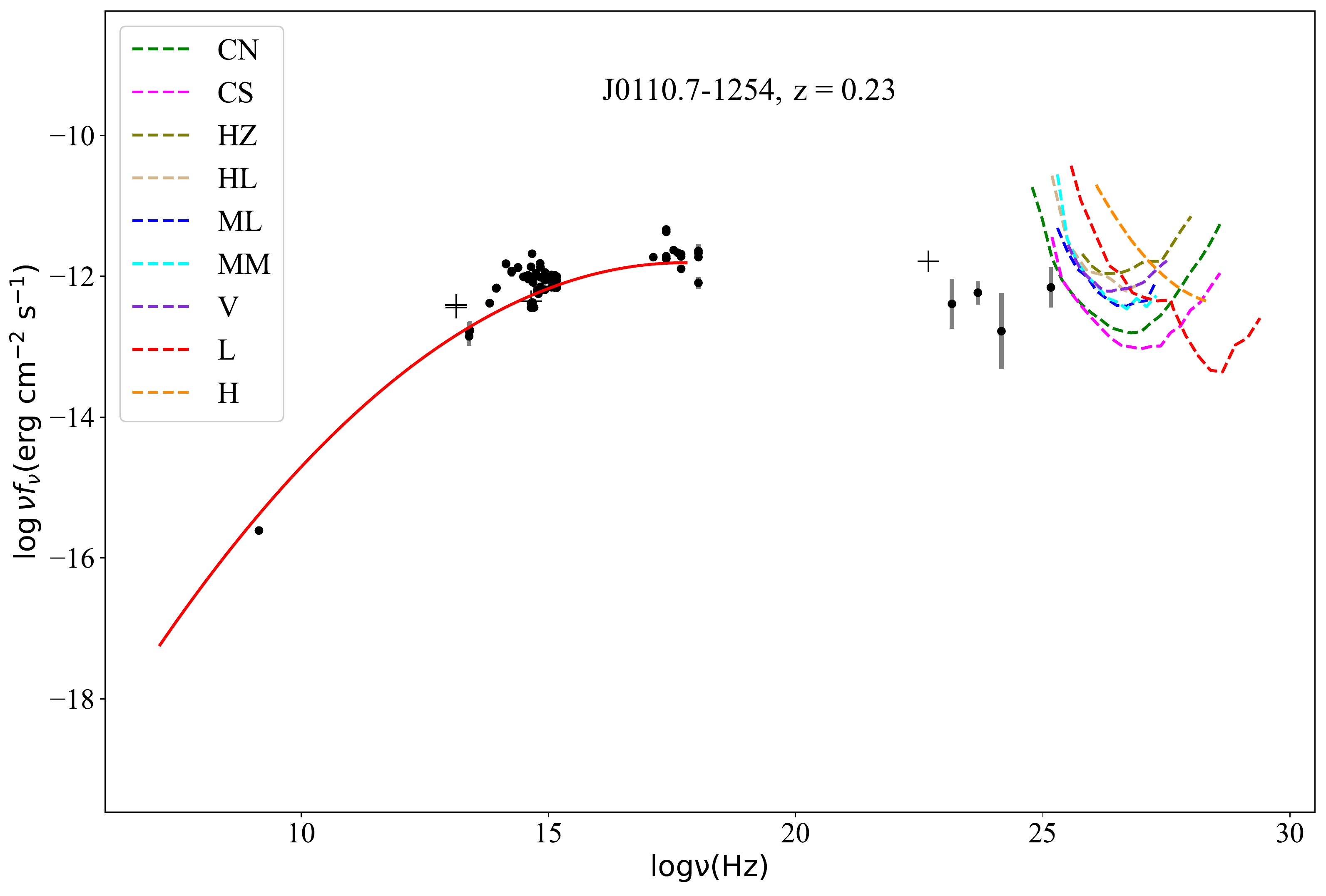}
\vspace{0.2in}
\includegraphics[width = 3.5 in]{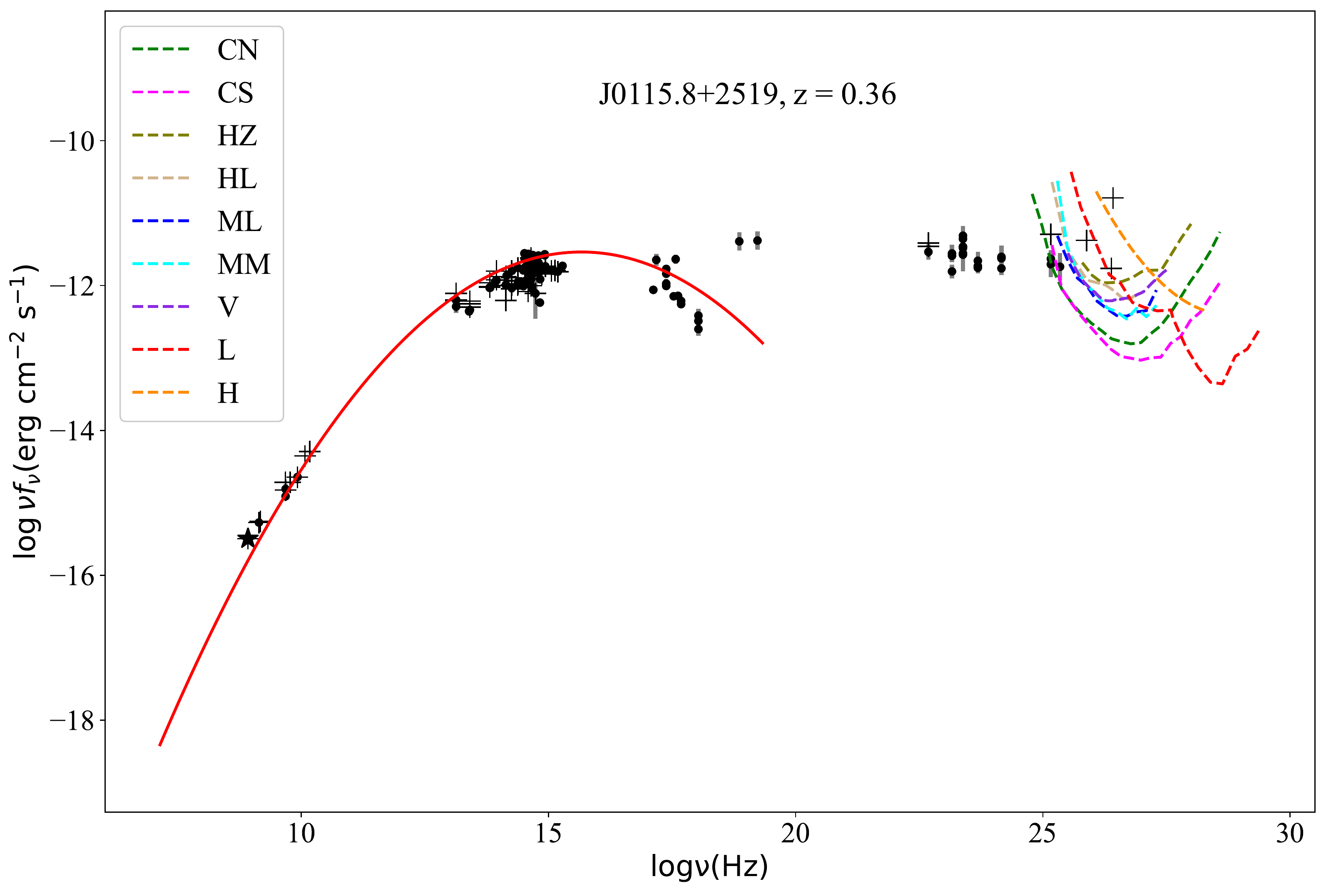}
\vspace{0.2in}
\includegraphics[width = 3.5 in]{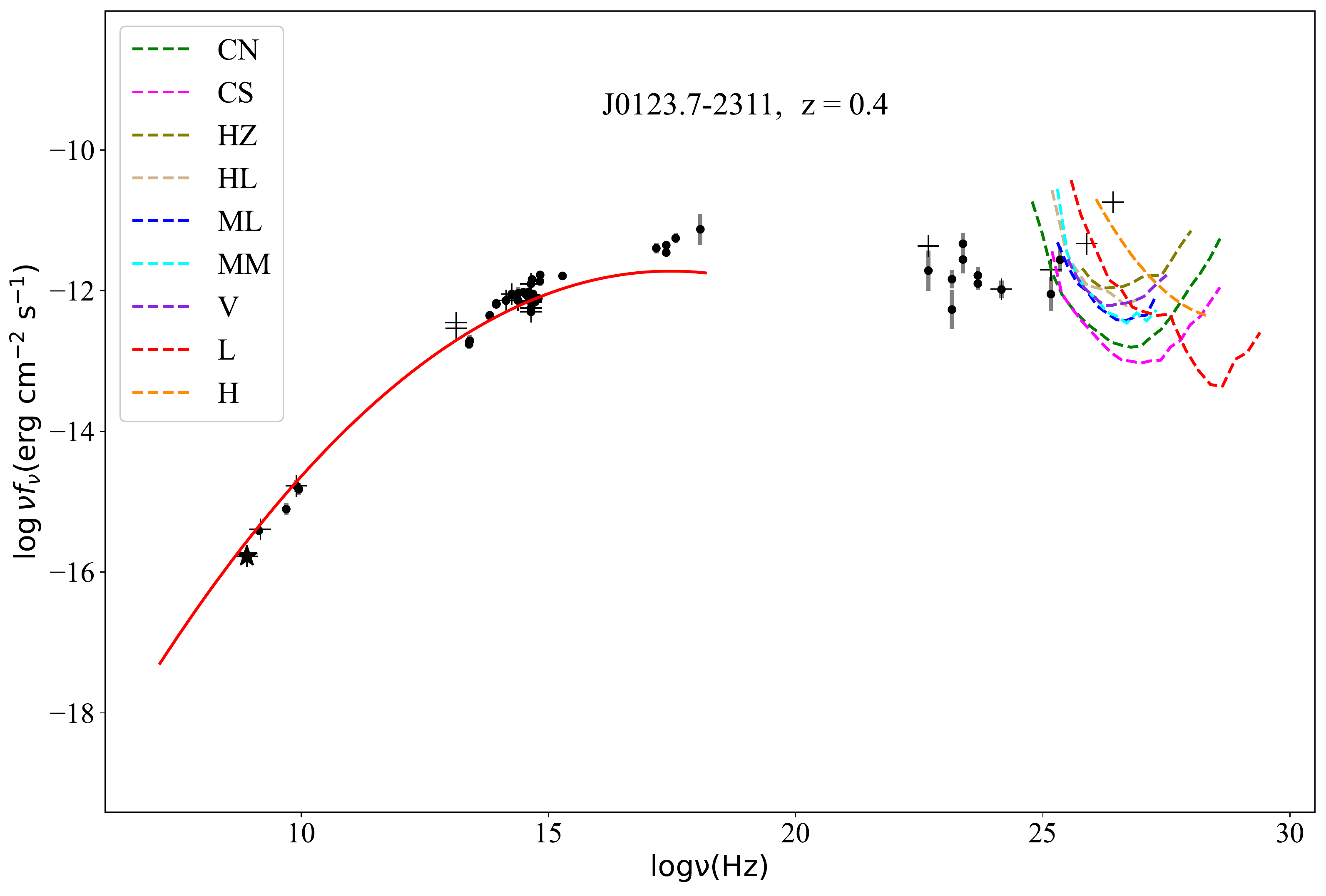}

\caption{SEDs for 40 TeV candidates. Only six items are displayed. A complete listing of this figure is available in the online version. Grey error bars stand for data points. Note that the TeV band photons are not certificated by IACTs and EAS arrays. Red and solid blue curves stand for optimal-fitting synchrotron bump and IC bump, while black dash-dot curves are EBL-absorbed IC bumps.
Three vertical dash-dot curves are $log \rm\nu$ (Hz) correndsponding to 1 TeV, 10 TeV, and 100 TeV. Green, magenta, olive, tan, blue, cyan, blue-violet, red, dark-orange dotted curves represent the sensitivity of CTAO north (north site, ${\rm 0}^\circ \leq Z \leq {\rm 20}^\circ$), CTAO south (south site, ${\rm 0}^\circ \leq Z \leq {\rm 20}^\circ$),
H.E.S.S zenith (${\rm 0}^\circ \leq Z \leq {\rm 5}^\circ$), H.E.S.S low (${\rm 12}^\circ \leq Z \leq {\rm 22}^\circ$), MAGIC low (${\rm 0}^\circ \leq Z \leq {\rm 30}^\circ$), MAGIC medium (${\rm 30}^\circ \leq Z \leq {\rm 45}^\circ$), VERITAS (${\rm 0}^\circ \leq Z \leq {\rm 20}^\circ$), LHAASO, and HAWC. For each source redshifts ($z$) are also shown.
Besides, the pentagon stands for data whose flux error is bigger than flux upper limits, and the cross is data in the low radio energy range (${\rm log} \nu (\:\mathrm{Hz}) < 9$).}
\label{sed}
\end{figure*}

The tactic is to correct the data by considering EBL first and then fitting SEDs on the intrinsic data.
The SED contains two bumps, a low-energy one in the infrared to soft X-ray energy range due to synchrotron emission and a high-energy one in the region between hard X-ray to $\gamma$-ray that is associated with IC emission. To do so, we use an online tool \footnote{http://tools.asdc.asi.it/SED/} provided by the space science data center (SSDC) of the Italian Space Agency to collect data from radio to TeV from many missions and experiments together with catalogs and archival data
\citep{2001A&A...366...62A,
2002babs.conf...63G,
2003ApJ...598..242A,
2003A&A...406L...9A,
2009ApJ...696L.150A,
2005ApJ...621..181D,
2005ApJ...634..947S,
2008ApJ...684L..73A,
2011ApJ...738..169A,
2008JPhG...35f5202G,
2010JPhG...37l5201C,
2012JPhG...39d5201C,
2010A&A...520A..83H,
2013MNRAS.434.1889H,
2011ApJ...742..127A,
2015ApJ...799....7A,
2011ApJ...734..110B,
2012ApJ...758....2B,
2013ApJ...776...69A,
2014ApJ...785L..16A,
2013ApJ...762...92A,
2013PhRvD..88j2003A,
2015ApJ...802...65A,
2015ApJ...812...60B,
2015NIMPA.770...42S}.
When we fit the SEDs, we discard the data bins whose flux error is bigger than flux upper limits. Note that there are TeV photons with zero error of the TeV candidates. Those TeV band photons are not certified by IACTs or EAS array (See Fig.~\ref{IC_fitting}). Likewise, we also discard data in the low radio energy range (${\rm log} \nu (\:\mathrm{Hz}) < 9$), since there is an asymmetry in the synchrotron bump of some sources.
In this paper, we do not take into account the simultaneous observation of the data.
EBL photons interact with VHE photons via pair production, a process that induces an attenuation of the observed gamma-ray spectra from blazars at cosmological distances, which may have a large impact on SED in IC bump.
Assuming the EBL optical depth ($\tau\ (E, z)$) to depend on the photon energy ($E$) and redshift ($z$), \citet{ebl} provided a 2D grid of $\tau\ (E, z)$ \footnote{https://www.ucm.es/blazars/ebl}in the range of 0.001 TeV $\leq E \leq$ 100 TeV, and 0 $\leq z \leq$ 6.
Subsequently, the two bumps of the observed SED were fitted separately with a log-parabola as follows \citep{Fan2016}:
\begin{equation}
\label{sed_equa}
\log (\nu f_{\nu}) = c (\log \nu - \log \nu_{\rm p})^2 + \log \nu_{\rm p}f_{\nu_{\rm p}},
\end{equation}
where $|2c|$ is the spectral curvature,
$\log \nu_{\rm p}$ is the logarithm of the peak frequency,
$\log \nu_{\rm p}f_{\nu_{\rm p}}$ is the logarithm of peak flux,
and $\log (\nu f_{\nu})$ is the flux density.

we successfully fit the synchrotron bump for 40 TeV candidates and the IC bump for 20 TeV candidates and return the errors of the fitting $|2c|$, $\log \nu_{\rm p}$ (in units of $\mathrm{Hz}$) and $\log \nu_{\rm p}f_{\nu_{\rm p}}$ (in units of $\mathrm{erg\ cm^{-2}\ s^{-1}}$).
The fitting process executes on \emph{scipy.optimize.curve\_fit} of \textsf{python}. The results are also listed in Tab.~\ref{IC_fitting},
in which Col. (1) gives the 4FGL name,
Col. (2) to Col. (4) are parameters and errors for fitting synchrotron bump,
Col. (5) to Col. (7) are parameters and errors for fitting IC bump,
while the right side of the slash is the data without EBL-absorbed.
For each high-confidence TeV candidate, the SEDs fitting results are shown in Fig.~\ref{sed}.

\begin{table*}[htbp]
\centering
\scriptsize
\caption{SEDs fitting results in the observed frame for 40 TeV candidates.$^{*}$}
\label{IC_fitting}
\scalebox{0.9}{
\begin{threeparttable}
\begin{tabular}{cccccccccccccc}
\midrule
\midrule
4FGL name &$\rm |c^s|$ &$\log\nu^{s}_{p}$ & $ {\log v^{s}_{p}}{f^{\rm s}_{\rm p}}$ & $\rm |c^{IC}|$ &$\rm {log\nu^{IC}_{p}}$ & ${\log v^{IC}_{p}}{f^{\rm IC}_{\rm p}}$ \\
(1)  & (2)  & (3)  & (4) & (5)  & (6)  & (7)\\
\midrule

J0037.8+1239 & 0.11 ± 0.02 & 15.35 ± 0.46 & -11.17 ± 0.07 & \multicolumn{1}{l}{0.02 ± 0.02} & \multicolumn{1}{l}{23.15 ± 2.42} & \multicolumn{1}{l}{-11.74 ± 0.1} \\
J0051.2-6242 & 0.16 ± 0.01 & 15.72 ± 0.06 & -11.23 ± 0.03 & \multicolumn{1}{l}{0.08 ± 0.07} & \multicolumn{1}{l}{25.57 ± 1.2} & \multicolumn{1}{l}{-11.37 ± 0.13} \\
J0110.1+6805 & 0.1 ± 0.01 & 14.99 ± 0.29 & -10.89 ± 0.09 &       &       &  \\
J0110.7-1254 & 0.05 ± 0.01 & 17.76 ± 1.09 & -11.81 ± 0.22 &       &       &  \\
J0115.8+2519 & 0.09 ± 0.01 & 15.67 ± 0.26 & -11.54 ± 0.04 &       &       &  \\
J0123.7-2311 & 0.05 ± 0.01 & 17.47 ± 0.88 & -11.72 ± 0.19 &       &       &  \\
J0159.5+1046 & 0.09 ± 0.0 & 15.63 ± 0.18 & -11.66 ± 0.03 & \multicolumn{1}{l}{0.03 ± 0.02} & \multicolumn{1}{l}{22.26 ± 0.67} & \multicolumn{1}{l}{-11.6 ± 0.11} \\
J0209.3-5228 & 0.11 ± 0.02 & 15.8 ± 0.09 & -11.08 ± 0.05 & \multicolumn{1}{l}{0.07 ± 0.04} & \multicolumn{1}{l}{24.99 ± 0.57} & \multicolumn{1}{l}{-11.43 ± 0.04} \\
J0211.2+1051 & 0.12 ± 0.0 & 14.8 ± 0.1 & -10.57 ± 0.02 & \multicolumn{1}{l}{0.06 ± 0.05} & \multicolumn{1}{l}{22.53 ± 1.04} & \multicolumn{1}{l}{-11.04 ± 0.07} \\
J0244.6-5819 & 0.12 ± 0.03 & 16.17 ± 0.49 & -11.15 ± 0.12 &       &       &  \\
...&...&...&...&...&...&...&\\ 
...&...&...&...&...&...&...&\\
...&...&...&...&...&...&...&\\
\midrule
\bottomrule
\end{tabular}
\begin{tablenotes}
\footnotesize
\item[*] Only ten items are displayed. A complete listing of this table is available in the online version.
\end{tablenotes}
\end{threeparttable}}
\end{table*}

\subsection{Visibility of TeV blazar candidates}
Being different from the all-sky scanning characteristics of space detectors, ground detectors can only scan part of the sky due to location constraints. 
IACTs need a dark sky for observations, and the field of view (FoV) of the current IACTs is small (3$^\circ−$ 5 $^\circ$). Besides, IACTs duty-cycle is restricted by the need to observe only during clear-sky, moonless nights, and, in addition, further constraints stem from the zenith angle: while the zenith angle increases, the sensitivity worsens and the energy threshold increases. This results in limiting the portion of the sky available for observation.
At the same time, CTAO, as the next generation of IACT, consists of two arrays located in the Northern (28$^\circ$ 45$^\prime$ N) and the Southern (24 $^\circ$ 41$^\prime$ S) hemispheres so that the FoV covers most of the sky, but not all, since CTAO suffers the same limitations affecting the current generation IACTs observations at high Zenith Angles ($Z$).
Conversely, the EAS arrays can work under all weather conditions and have large FoV ($\sim {\rm 2} sr$), which can continuously monitor a significant fraction of the sky every day.

We compile the sensitivity curves of IACTs and EAS arrays from \cite{HL, MAGIC_sensitivity, LHAASO, HAWC_sensitivity, HAWC_region} and online database: CTAO\footnote{https://www.cta-observatory.org/science/ctao-performance/}, H.E.S.S.
\footnote{chrome-extension://efaidnbmnnnibpcajpcglclefindmkaj/https://www.mpi-hd.mpg.de/hfm/HESS/pages/home/proposals/sc\_sens.pdf}, VERITAS\footnote{https://veritas.sao.arizona.edu/about-veritas/veritas-specifications.}.
The FoV of these sensitivity curves are as follows: 
CTAO north (north site and $\mathrm{0}^\circ \leq Z \leq \mathrm{20}^\circ$), CTAO south (south site and $\mathrm{0}^\circ \leq Z \leq \mathrm{20}^\circ$),
H.E.S.S. zenith ($Z \sim \mathrm{0}^\circ$), H.E.S.S. medium ($\mathrm{12}^\circ \leq Z \leq \mathrm{22}^\circ$), MAGIC low ($\mathrm{0}^\circ \leq Z \leq \mathrm{30}^\circ$), MAGIC medium ($\mathrm{30}^\circ \leq Z \leq \mathrm{45}^\circ$), VERITAS ($\mathrm{0}^\circ \leq Z \leq \mathrm{20}^\circ$), LHAASO ($Z \leq \mathrm{40}^\circ$ or $\mathrm{-11}^\circ \leq Dec \leq \mathrm{69}^\circ$), and HAWC ($\mathrm{-20}^\circ \leq Dec \leq \mathrm{60}^\circ$); while the exposure times are: 507 days for HAWC, one year for LHAASO, 25 hours for H.E.S.S. near the zenith, and 50 hours for the rests.

\begin{figure*}[htbp]
\centering
\includegraphics[width = 2.5 in]{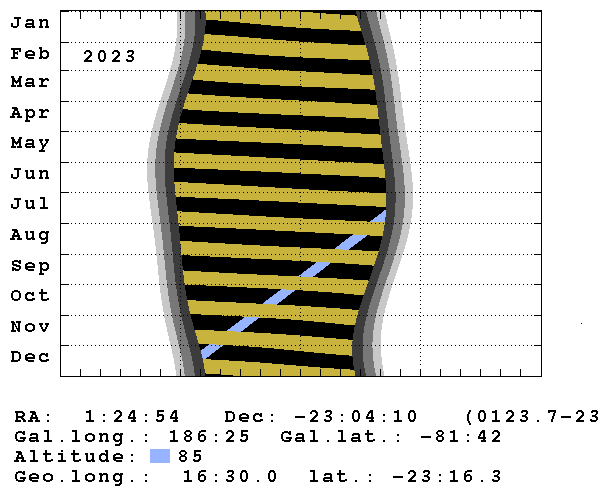}
\hspace{0.1 in}
\vspace{1 in}
\includegraphics[width = 2.5 in]{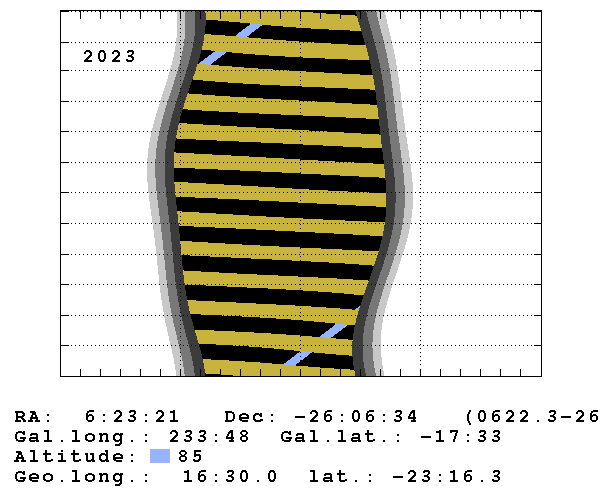}
\caption{Four sources that could pass by the observation window of H.E.S.S near the zenith in the whole year of 2023. They are 4FGL J0123.7-2311, 4FGL J0622.3-2605.
A complete display of the visibility of the source for other IACTs and EAS arrays is available in the supplementary data section of the online version.
The figures are generated by H.E.S.S. online visibility tool.
The sun is up during the times indicated by the white areas. Grey levels correspond to civil, naval, and astronomical twilight, respectively.
The moon is up or making twilight in the yellow areas.
The blue colors indicate the times when the object is above given altitudes.
RA/Dec positions listed below the plot are for the current epoch, not J2000.
}
\label{HESSTOOL_observable}
\end{figure*}

Since EAS arrays can observe all sources in the FoV simultaneously according to the source declination, whereas IACTs can only track one source at a time and require darkness sky,
we employ two different strategies to infer the visibility of a given source.
A source could pass the observation window for EAS arrays if its declination is within the most observable sky.
On the other hand, for IACTs, we evaluate observation windows by 2 tools: the H.E.S.S. online visibility tool \footnote{https://www.mpi-hd.mpg.de/hfm/HESS/pages/home/visibility/} and 
a \textsf{python} package: \textsf{astroplan}.
Input the celestial coordinates of the source and the observation information of IACTs to these two tools. We can get visible months during darkness at a specific elevation angle (1 - $Z$), as shown in Fig.~\ref{HESSTOOL_observable} and Tab.~\ref{months_observable}.
The 2 tools have slight differences in the specific observable months. 
The observable months of 40 high-confidence TeV blazar candidates for the 2 tools are also shown in Tab.~\ref{months_observable},
where
Col. (1) gives the source name;
Col. (2) and Col. (7) give the observable months for IACTs and EAS arrays;
The numbers represent the observable months, i.e. `1' for January, `2' for February, and so on.

\subsection{High-confidence TeV blazar candidates}
We calculate the $\tau\ (E, z)$ for 40 TeV candidates with $E$ in the range of 0.001TeV to 100 TeV, and get the EBL-absorbed IC bumps.
By comparing the 20 IC bumps and the sensitivity curves of IACTs and EAS arrays in the energy range of 1 TeV $\leq E \leq$ 100 TeV, we obtain 7 out of 40 high-confidence TeV blazar candidates that satisfy the two detectable criteria. The source number can be detected by the TeV facilities are (see Tab.~\ref{IACTs} for detail): 6 CTAO (north or south) targets, 2 H.E.S.S targets, 5 MAGIC targets, and 1 VERITAS targets, and EAS arrays: 1 LHAASO targets, 0 HAWC targets.
The SML and detectability results of the 40 high-confidence TeV blazar candidates are shown in Tab.~\ref{IACTs}, where
Col. (1) gives the 4FGL name;
Col. (2) and Col. (3) give the galactic coordinates (in degrees);
Col. (4) synchrotron-peak frequency ($\log\nu^{s}_{p}$);
Col. (5) redshift (z);
Col. (6) flux density EBL-absorbed of sources at $1\:\mathrm{TeV}$ ($f_{1\:\mathrm{TeV}}$) which can be computed using Formula~(\ref{sed_equa}) the parameters of the IC bump;
Col. (7) to Col. (10) $logistic$ in 4 catalogs, which is the likelihood of a source belonging to TeV sources given its optimal features 
Col. (11) the SED class in 4FGL-DR2 / 4LAC-DR2, where `bll' stands for BL Lac, `fsrq' for FSRQ, and `bcu' for blazar candidate of uncertain type;
Col. (12) to Col. (14) candidates compared with those of 3FHL, 3HSP, and 2BIGB,
a candidate in common is marked as `Y';
Col. (15) a candidate comparing with those of other literature, `C \& G' represents the candidate is the same with
\citet{2002AA...384...56C},
while `M' for
\citet{2013ApJS..207...16M},
`F' for
\citet{2019MNRAS.486.1741F}
`Chi' for
\cite{Chiaro_2019};
Col. (16) IC bump of candidates compare with the sensitivity of IACTs and particle detectors arrays in the range of $\mathrm{1 TeV} \leq E \leq \mathrm{100 TeV}$.
`CN', `CS',  `HZ', `HL', `ML', `MM', `V', `L,', and `H', stand for CTAO north, CTAO south,
H.E.S.S zenith, H.E.S.S low, MAGIC low, MAGIC medium, VERITAS, LHAASO, and HAWC, respectively.

We also want to know how far away the source becomes undetectable in the TeV energy band, i.e. the redshift upper limit of the source beyond which it will become undetectable. For this purpose, in the range of 0 $\leq z \leq$ 6, we compare the EBL-absorbed IC bumps for 20 blazars and the sensitivity curves of detectors, at $E = $ 1 TeV, 10TeV, and 100 TeV, respectively. Redshifts upper limits for the 20 blazars for each IACT and EAS array are listed in Tab.~\ref{z_limits}, where
Col. (1) gives the 4FGL name;
Col. (2) to Col. (10) are the redshift upper limits of the 20 blazars for each, where the values in each column are redshift upper limits at $E = $1 TeV, $E = $10 TeV, and $E = $100 TeV, respectively. The value `0' means the IC bump can not be detected at the specific $E$;
Col. (11) redshifts;
Col. (12) is the detectability for Cherenkov detectors.
One can see the VHE photons could survive from EBL easier than those with higher redshift. This can also be seen from Formula~(\ref{logit1}) $\sim$ (\ref{logit4}),
the negative coefficient of $z$ leads $logit$ to be inversely proportional to $z$.
A larger redshift will result in a smaller logistic, indicating a source to be less likely to be a TeV candidate.

\begin{table}[htbp]
\scriptsize
\centering
\caption{Observable months of IACTs and EAS arryas$^{*}$}
\label{months_observable}
\scalebox{0.9}{
\begin{threeparttable}
\begin{tabular}{cccccccc}
\toprule
\toprule
4FGL name & CTAO North & CTAO South & H.E.S.S. zenith & H.E.S.S. low & MAGIC low & MAGIC medium & VERITAS \\
(1)  & (2)  & (3)  & (4) & (5)  & (6)  & (7) & (8)\\
\midrule
J0037.8+1239 & 7, 8, ..., 12 &       &       &       & 1, 7, ..., 12 & 1, 6, ..., 12 & 7, 8, ..., 12 \\
J0051.2-6242 &       &       &       &       &       &       &  \\
J0110.1+6805 &       &       &       &       &       & 1, 7, ..., 12 &  \\
J0110.7-1254 &       & 7, 8, \ldots, 12 &       & 6\tnote{h},  7, ..., 12 &       & 1, 7, ..., 12 &  \\
J0115.8+2519 & 1, 7, ..., 12 &       &       &       & 1, 7, ..., 12 & 1, 2, 6\tnote{h}, ..., 12 & 1, 7\tnote{a}, ..., 12 \\
J0123.7-2311 &       & 7, 8, ..., 12 & 7, 8, ..., 12 & 6\tnote{h}, 7, ..., 12 &       &       &  \\
J0159.5+1046 & 1, 8, ..., 12 &       &       &       & 1, 7, ..., 12 & 1, 2, ..., 12 &  \\
J0209.3-5228 &       &       &       &       &       &       &  \\
 J0211.2+1051 & 1, 8, ..., 12 &       &       &       & 1, 8, ..., 12 & 1, 2, 7, ..., 12 &  \\
J0244.6-5819 &       &       &       &       &       &       &  \\
...&...&...&...&...&...&...&...\\
...&...&...&...&...&...&...&...\\
...&...&...&...&...&...&...&...\\
\midrule
\end{tabular}
\begin{tablenotes}
\footnotesize
\item[*] Only ten items are displayed. A complete listing of this table is available in the online version.
\item[h] `h' indicates the month is only visible for the H.E.S.S. tool. 
\item[a] `a' indicates the month is only visible for \textsf{astroplan}.
\end{tablenotes}
\end{threeparttable}}
\end{table}

\begin{sidewaystable}[htbp]
\centering
\caption{Predicting results for 40 TeV blazar candidates$^{*}$}
\label{IACTs}
\scalebox{0.85}{
\begin{threeparttable}
\begin{tabular}{cccccrcccccccccc}
\midrule
\midrule
\cmidrule{1-5}\cmidrule{7-15}    \multirow{2}[4]{*}{4FGL name} & \multirow{2}[4]{*}{GLON} & \multirow{2}[4]{*}{GLAT} & \multirow{2}[4]{*}{$\log \nu^{\rm s}_{\rm p}$} & \multirow{2}[4]{*}{$z$} & \multirow{2}[4]{*}{$\log f_{\rm 1 TeV}$} & \multicolumn{4}{c}{$\ logistic$}  & \multirow{2}[4]{*}{Class} & \multicolumn{3}{c}{Other catalogs} & \multirow{2}[4]{*}{Common} & \multirow{2}[4]{*}{Detectability} \\
\cmidrule{7-10}\cmidrule{12-14}          &       &       &       &       &       & 4FGL  & 3FHL  & 3HSP  & 2BIGB &       & 3FHL  & 3HSP  & 2BIGB &       &  \\
(1) & (2) & (3) & (4) & (5) & (6) & (7) & (8) & (9) & (10) & (11) & (12) & (13) & (14) & (15) & (16)\\
\midrule
J0037.8+1239 & 117.77 & -50.08 & 15.05 & 0.09  & -12.32 & 82.50\% & 54.60\% &       &       & bll   & Y     &       &       &       & CN, CS, ML, MM \\
J0051.2-6242 & 302.96 & -54.43 & 15.96 & 0.3   & -12.91 & 84.10\% &       &       &       & bll   &       &       &       &       &  \\
J0110.1+6805 & 124.69 & 5.29  & 14.85 & 0.29  &  & 81.80\% &       &       &       & bll   &       &       &       &       &  \\
J0110.7-1254 & 141.53 & -75.08 & 17    & 0.23  &       & 82.40\% &       &       &       & bll   &       &       &       &       &  \\
J0115.8+2519 & 129.85 & -37.21 & 15.75 & 0.36  &  & 97.60\% &       &       &       & bll   &       &       &       &       &  \\
J0123.7-2311 & 186.41 & -81.7 & 17.96 & 0.4   &  & 91.50\% &       &       &       & bll   &       &       &       &       &  \\
J0159.5+1046 & 148.75 & -48.66 & 15.8  & 0.2   & -13.04 & 87.70\% & 53.60\% &       &       & bll   & Y     &       &       &       &  \\
J0209.3-5228 & 278.35 & -60.77 & 16.14 & 0.16  & -12.28 & 95.60\% &       & 67.70\% & 60.20\% & bll   &       & Y     & Y     & M     &  \\
J0211.2+1051 & 152.59 & -47.37 & 14.22 & 0.2   & -12.86 & 88.90\% & 55.50\% &       &       & bll   & Y     &       &       &       &  \\
J0244.6-5819 & 278.45 & -53.09 & 17.03 & 0.27  &       & 88.00\% & 53.30\% &       & 48.50\% & bll   & Y     &       & Y     & F     &  \\
...&...&...&...&...&...&...&...&...&...&...&...&...&...&...&...\\
...&...&...&...&...&...&...&...&...&...&...&...&...&...&...&...\\
...&...&...&...&...&...&...&...&...&...&...&...&...&...&...&...\\
\midrule
\end{tabular}
\begin{tablenotes}
\footnotesize
\item[*] Only ten sources are displayed. A complete listing of this table is available in the online version.
\item[**] The source can be detected in $\mathrm{1 TeV} \leq E \leq \mathrm{10 TeV}$ and $\mathrm{10 TeV} \leq E \leq \mathrm{100 TeV}$
\end{tablenotes}
\end{threeparttable}}
\end{sidewaystable}%

\begin{table*}[htbp]
\centering
\scriptsize
\caption{Redshift uplimit of the 20 blazars for each IACT and EAS array.$^{*}$}
\label{z_limits}
\scalebox{0.8}{
\begin{threeparttable}
\begin{tabular}{cccccccccccc}
\midrule
\midrule
4FGL name & $z_{\rm uplimit}$ of CN&$z_{\rm uplimit}$ of CS & $z_{\rm uplimit}$ of HZ & $z_{\rm uplimit}$ of HL & $z_{\rm uplimit}$ of ML & $z_{\rm uplimit}$ of MM & $z_{\rm uplimit}$ of V & $z_{\rm uplimit}$ of L & $z_{\rm uplimit}$ of H & $z$ & Detectability\\
(1)   & (2)   & (3)   & (4)   & (5)   & (6)   & (7)   & (8)   & (9)   & (10) & (11) & (12)\\
\midrule
J0037.8+1239 &  & 0.2, 0.05, 0.01 &  & 0.02, 0.01, 0.01 &  & 0.09, 0.01, 0.01 &  & 0, 0.01, 0.01 & 0, 0, 0.01 & 0.09  & \multicolumn{1}{l}{CN, CS, ML, MM} \\
J0051.2-6242 &  & 0.29, 0.07, 0.01 & & 0.14, 0.03, 0.01 & 0.20, 0.02, 0.01 & 0.19, 0.04, 0.01 & 0.17, 0.01, 0.01 & 0.1, 0.04, 0.01 & 0, 0.02, 0.01 &  0.3     &  \\
J0110.1+6805 &  &  &  &  &  & & &  &  &    0.29   &  \\
J0110.7-1254 &  &  &  &  &  &  & &  &  &    0.23   &  \\
J0115.8+2519 &  &  &  &  &  &  &  &  &  &   0.36    &  \\
J0123.7-2311 & &  &  &  &  &  &  &  & &   0.4    &  \\
J0159.5+1046 &  &  &  & & &  &  & 0, 0, 0.01 &  &   0.2    &  \\
J0209.3-5228 &  & 0.27, 0.07, 0.01 &  & 0.11, 0.02, 0.01 & 0.17, 0.01, 0.01 & 0.17, 0.02, 0.01 &  & 0.07, 0.03, 0.01 & 0, 0.01, 0.01 &   0.16    &  \\
J0211.2+1051 &  &  &  &  &  &  &  & 0, 0, 0.01 &  &   0.2    &  \\
J0244.6-5819 &  &  &  &  &  &  & &  &  &    0.27   &  \\
...&...&...&...&...&...&...&...&...&...&...&...\\
...&...&...&...&...&...&...&...&...&...&...&...\\
...&...&...&...&...&...&...&...&...&...&...&...\\
\bottomrule
\end{tabular}
\begin{tablenotes}
\footnotesize
\item[*] Only ten items are displayed. A complete listing of this table is available in the online version.
\end{tablenotes}
\end{threeparttable}}
\end{table*}

\section{Discussions and conclusions}
\subsection{Discussions}
The SFS identified five features in 4FGL-DR2 / 4LAC-DR2 that are most critical to distinguish TeVs from non-TeVs: flux density, FOM, synchrotron peak frequency, redshift, spectral index, and variability, which are compatible with \citet{Lin2016}. Further considering X-ray emission and the $\log \nu_{\rm p}^{\rm s}$ shifts in behavior during a flare should be further analyzed in future work, sinces blazars that are not generally detected at TeV energies could be detected during flaring states.
Therefore, it is possible that all blazars can emit the TeV photons.
The blazars with TeV emission detected are only because their flux density in the TeV energy band is relatively high in specific periods, such as the flaring period.
Note that 4FGL-DR2 includes features such as $Variability\_Index$, which mirrored the difference between the flux fitted in each time interval and the average flux over the entire catalog interval.
Furthermore, $Frac\_Variability$ is the fractional variability computed from the fluxes in each year. The $logit^{\prime}$ of 4FGL-DR2 / 4LAC-DR2 contains the $Frac\_Variability$ making sure the LR is sensitive to the sources with synchrotron peak shifting during flaring states.

\begin{figure}
\centering
\includegraphics[width = 3.5 in]{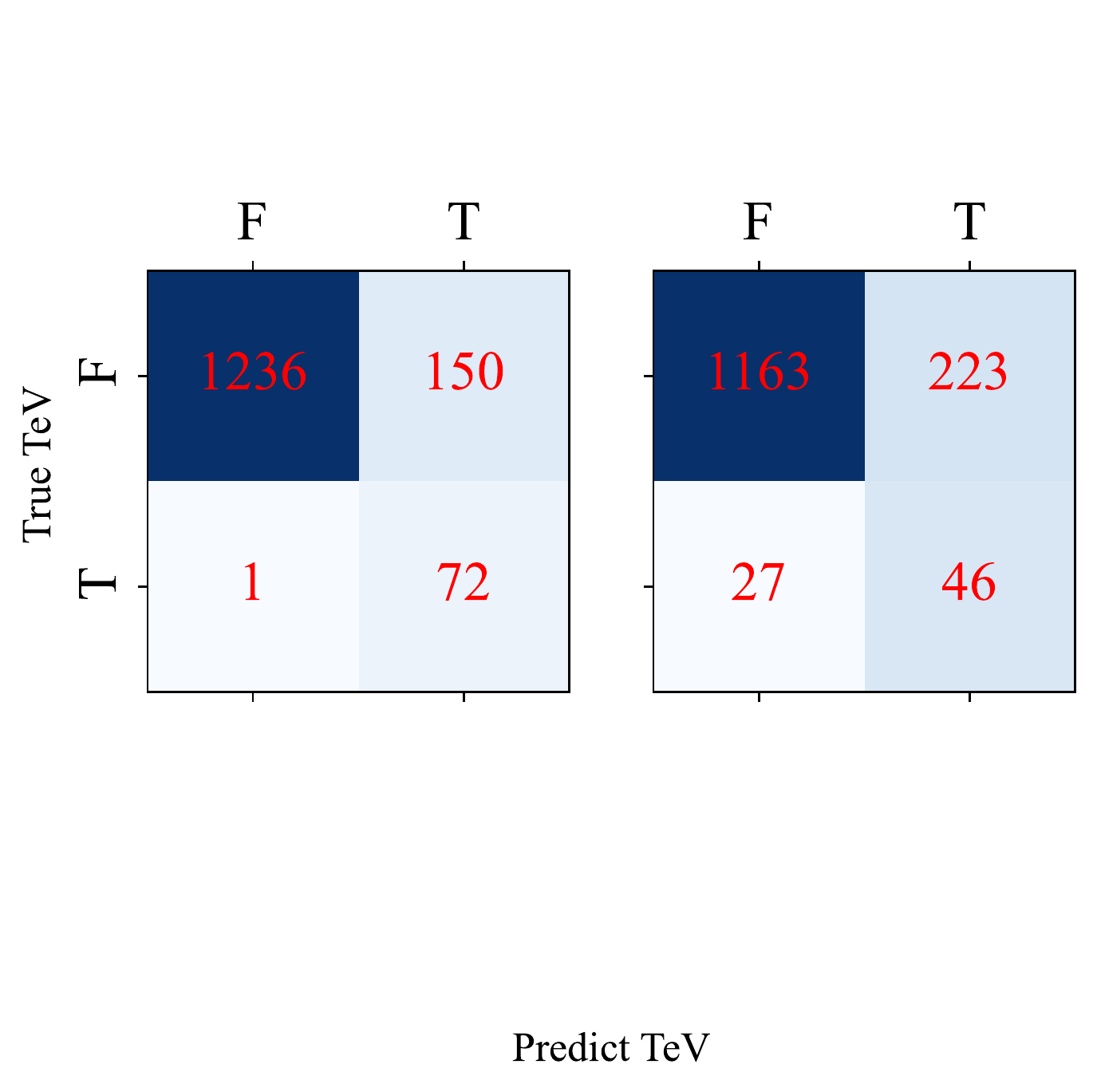}
\caption{
Confusion matrix for SML and HSPs,
where the coordinate value `T' stands for `True', and `F' stands for `False'.
Left panel: Confusion matrix for 4FGL-DR2 / 4LAD-DR2, with 5 FOFs:
$\Gamma$
$V_{F}$
$\log f ^{\rm ph}_{\rm 7}$
$z$.
Right panel: Confusion matrix for $\log \nu_{\rm p}^{\rm s} \geqslant$ 15.3 Hz.
True positives (top left),
false negatives (top right),
false positives (bottom left),
and true negatives (bottom right)
for both panels;
}
\label{HSP_cmat_4FGL_4LAC}
\end{figure}

\begin{figure}
\centering
\includegraphics[width = 3.5 in]{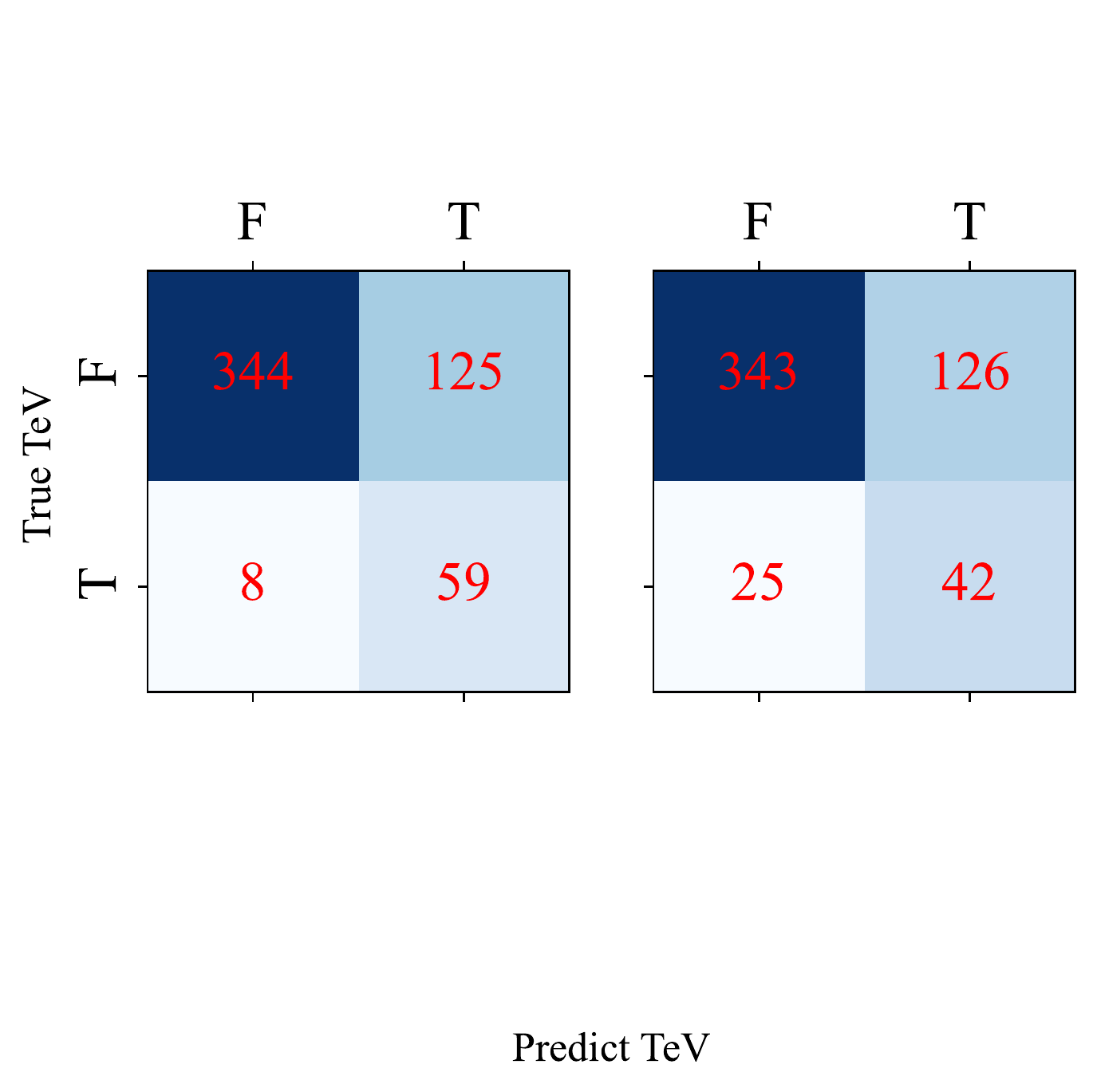}
\caption{
Confusion matrix for SML and HSPs,
where the coordinate value `T' stands for `True', and `F' stands for `False'.
Left panel: Confusion matrix for 3FHL, with 2 FOFs:
$\log f ^{\rm ph}_{\rm 2}$,
$z$.
Right panel: Confusion matrix for $\log \nu_{\rm p}^{\rm s} \geqslant$ 15.3 Hz.
True positives (top left),
false negatives (top right),
false positives (bottom left),
and true negatives (bottom right)
for both panels.
}
\label{HSP_cmat_3FHL}
\end{figure}

According to the classification proposed by
\citet{Abdo2010} or
\citet{Fan2016},
sources with $\log \nu_{\rm p}^{\rm s} >$ 15 (or $\log \nu_{\rm p}^{\rm s} >$ 15.3) are HSPs.
In TeVCat, if we considered only HSP/HSP BL Lacs, we would have ignored more than 11\% of TeV sources.
We presented here an alternative way, where we have not applied many initial cuts by filtering on SED class or other properties.
To compare the performance of two criteria: HSPs ($\log \nu_{\rm p}^{\rm s}\geqslant 15.3$ Hz \citep{Fan2016}) and LR model, we calculate the metric and confusion matrix on the testing sets of 4FGL-DR2 / 4LAC-DR2 and 3FHL.
For 4FGL-DR2 / 4LAC-DR2, the $AUC$: 73\% and $TPR$: 62\% are both lower than those predicted by LR, up to 27 TeV blazars are misclassified as non-TeVs.
For 3FHL, take off four sources without $\log \nu_{\rm p}^{\rm s}$ data, the $AUC$: 68\% and $TPR$: 63\% are also lower than those predicted by LR,
up to 25 TeV blazars are misclassified as non-TeVs.
The relevant confusion matrix is shown in Fig.~\ref{HSP_cmat_4FGL_4LAC} and Fig.~\ref{HSP_cmat_3FHL}. 
The results illustrated that our selection criterion is more effective.

So far,
TeV candidate selection techniques privileged targets with high
predicted TeV fluxes. In our analysis, instead, we identified targets
that, according to their broadband features, have high chances to become promising TeV targets in specific activity states and that may
have been missed so far due to their current status or to the lack of
the proper instruments and observing strategies. As a data-driven algorithm, ML is inevitably sensitive to the sample composition, the ratio of the training set and test set, the value of \textit{k} in cross-validation, etc.
Our result is to ensure its generalization based on existing data.
There is a potential bias in the fact that a significant fraction of TeVCat sources are Fermi-LAT detected blazars, as Fermi-LAT catalogs are browsed to list potential candidates for IACTs, and because Fermi-LAT monitoring is used to trigger Target of Opportunity programs.
Also, our observation proposal does not consider the weather or available time of detectors.
A feasible observation proposal should be more reliable and agreed upon by the observatory that considers more influencing factors.

Detecting extragalactic photons in the TeV band needs to face two major limitations, which are EBL and the performance of detectors (sensitivity, effective area, FoV, etc).
The relationship between the improvement of detector sensitivity and the detection of more TeV sources is far from simple, because if it is true that an improvement in sensitivity indeed helps detect more sources but, since the attenuation rapidly increases with energy and redshift, the overall effect is rather mild (see for example \citealt{2012MNRAS.422.3189G}).
The redshifts of the 81 blazars listed in TeVCat are taken from the four catalogs and literature, noting that for 4FGL J2243.9+2021 (or RGB J2243+203) we took the upper limit of the redshift range of 0.75$\leq z \leq$1.1 \citep{2019ApJ...884L..17S}. Then we calculate the $\tau\ (E, z)$ fixing $E = $ 1 TeV and we evaluate $\tau\ (E, z) \leq$ 13.1 for all 81 blazars. The results are also listed in Tab.~\ref{81tau_csv}, where
Col. (1) and Col. (2) give the TeVCat name and 4FGL-DR2 name;
Col. (3) and Col. (4) are the celestial equator coordinates (in degrees);
Col. (5) the redshift;
Col. (6) the $\tau\ (E, z)$ provided by \cite{ebl} at $E = $ 1 TeV.

\begin{table}[htbp]
\centering
\caption{$\tau\ (E, z)$ at $E = $1 TeV for the 81 TeV blazars$^{*}$}
\begin{tabular}{cccccc}
\toprule
\toprule
TeVCat Name & 4FGL Name & Ra & Dec & z     & $\tau$ \\
(1)   &  (2)   &  (3)   &  (4)   &  (5)   &  (6) \\
\toprule
J0013-188 & J0013.9-1854 & 3.47  & -18.89 & 0.095 & 0.92 \\
J0033-193 & J0033.5-1921 & 8.4   & -19.35 & 0.61  & 7.61 \\
J0035+598 & J0035.9+5950 & 8.82  & 59.79 & 0.467 & 5.7 \\
J0112+227 & J0112.1+2245 & 18.02 & 22.74 & 0.265 & 2.96 \\
J0136+391 & J0136.5+3906 & 24.14 & 39.1  & 0.75  & 9.35 \\
J0152+017 & J0152.6+0147 & 28.14 & 1.78  & 0.08  & 0.76 \\
J0214+517 & J0214.3+5145 & 33.57 & 51.75 & 0.049 & 0.45 \\
J0218+359 & J0221.1+3556 & 35.27 & 35.94 & 0.944 & 11.54 \\
J0222+430 & J0222.6+4302 & 35.67 & 43.04 & 0.444 & 5.39 \\
J0232+202 & J0232.8+2018 & 38.22 & 20.27 & 0.139 & 1.4 \\
...&...&...&...&...&...\\
...&...&...&...&...&...\\
...&...&...&...&...&...\\
\toprule
\end{tabular}%
\label{81tau_csv}%
\begin{tablenotes}
\item[*] Only ten items are displayed. A complete listing of this table is available in the online version.
\end{tablenotes}
\end{table}

All in all, the validation of TeV candidates is counting on IACTs and EAS arrays, while each of the IACTs and EAS arrays has advantages and disadvantages.
IACTs have better energy and angular resolution.
Besides, the energy threshold of IACTs could reach the GeV band while EAS arrays are generally sensitive above $\sim$ 10 TeV, but at $\sim$ 30 TeV, LHAASO sensitivity is comparable to that of CTAO, and LHAASO at $\sim$ 100 TeV is the most sensitive instrument \citep{LHAASO}.
On the other hand, EAS array observation time is not limited to moonless nights and has a much larger FoV, and the exposure time of IACTs is typically quoted for 50 hr whereas for 1 or 5 years of EAS arrays.
Although IACTs are better at capturing transient events, larger FoVs and longer exposure time make up the gap for the EAS array.
Overall, for IACTs and EAS arrays, the differences in their operation modes and performance parameters allow them to complement each other.
Furthermore, The sensitivity of LHAASO and the next-generation IACT array: CTAO, has been improved by an order of magnitude compared with the current IACTs and EAS arrays.
They have pushed the detection limit of the TeV band to higher than 100 TeV \citep{CTAO, LHAASO}, enabling us to detect the whole energy range of the TeV sky in the future.

\subsection{Conclusions}
In this paper,
we implement a machine learning algorithms called \emph{Logistic Regression} to search the TeV blazar candidates from 4FGL-DR2 / 4LAC-DR2, 3FHL, 3HSP, and 2BIGB. Furthermore, we filter out potential observation targets for IACTs and EAS arrays.
The main conclusions are enumerated below:
\begin{enumerate}
\item For the 4 catalogs, \emph{Logistic Regression} provides an empirical formula to find blazars that could be detected at TeV energies with $logit^{\prime} \geqslant \rm 0$.

For 4FGL-DR2 / 4LAC-DR2:
\begin{equation*}
logit^{\prime} = 4.808 \Gamma + 2.809 V_{F} + 3.889 \log f^{ph}_{\rm 7} - 3.34 z + 0.857 \log \nu_{\rm p}^{\rm s} + 20.244,
\end{equation*}
with $AUC =$ 98\%, and the $FPR =$ 11\%, $TPR =$ 99\%, and the $p_{\rm thre} =$ 40\%.
40 out of 150 non-TeV blazars have $logistic \geqslant$ 80\% and are thus expected to be high-confidence candidates.

For 3FHL:
\begin{equation*}
logit^{\prime} = 0.116 \log f ^{ph}_{\rm 2} - 0.628 z + 1.028,
\end{equation*}
with $AUC =$ 88\%, $FPR =$ 27\%, $TPR =$ 88\%, and the $p_{\rm thre =}$ 52\%.
24 out of 126 TeV candidates are common with the high-confidence ones of 4FGL-DR2 / 4LAC-DR2;

For 3HSP:
\begin{equation*}
logit^{\prime} = 0.306 FOM - 3.861 z - 0.173,
\end{equation*}
with $AUC =$ 98\%, $FPR =$ 2\%, $TPR =$ 91\%, and $p_{\rm thre =}$ 61\%.
11 out of 40 TeV candidates are common with the high-confidence ones of 4FGL-DR2 / 4LAC-DR2;

For 2BIGB,
\begin{equation*}
logit^{\prime} = -0.016 E_{\rm P} + 0.248 FOM - 4.395z - 0.122,
\end{equation*}
with $AUC =$ 97\%, $FPR =$ 8\%, $TPR =$ 96\%, and $p_{\rm thre =}$ 48\%.
14 out of 83 TeV candidates are common with the high-confidence ones of 4FGL-DR2 / 4LAC-DR2;

\item For the 40 high-confidence candidates,
we independently fit the two bumps of the SEDs,
infer the visibility in the sky.
For EAS arrays and IACTs in 2023, 7 candidates are potential targets:
6 CTAO targets, 2 H.E.S.S targets, 5 MAGIC targets, 1 VERITAS, 1 LHAASO targets, and 0 HAWC targets, respectively.

\item We get 2 common sources with \citet{2002AA...384...56C},
9 ones with \citet{2013ApJS..207...16M},
4 ones with \citet{2019MNRAS.486.1741F}, 
and none with \citet{2019ApJ...887..104C}, respectively.
\end{enumerate}

\section*{Acknowledgement}
Thanks are given to the excellent reviewer for the constructive comments and suggestions!
The work is partially supported by the National Natural Science Foundation of China
(NSFC U1831119, NSFC U2031201, NSFC 11733001, NSFC 12203034), Guangdong Major Project of Basic and Applied Basic Research (2019B030302001), Shanghai science and Technology Fund (22YF1431500), Scientific and Technological Cooperation Projects(2020-2023) between the People's Republic of China and the Republic of Bulgaria, Astrophysics Key Subjects of Guangdong Province and Guangzhou City,
the Large High Altitude Air Shower Observatory collaboration,
the Cherenkov Telescope Array,
the High Altitude Water Cherenkov Observatory,
the Major Atmospheric Gamma Imaging Cherenkov Telescopes,
the Very Energetic Radiation Imaging Telescope Array System,
and the UK Swift Science Data Centre at the University of Leicester.

\clearpage
\bibliographystyle{aasjournal}
\bibliography{ref}

\end{document}